\documentclass[11pt,preprint]{aastex}

\usepackage{graphicx}

\begin{document}

\title{Three-Dimensional Relativistic Magnetohydrodynamic Simulations of
Current-Driven Instability with A Sub-Alfv\'{e}nic Jet: Temporal Properties}

\author{
Yosuke Mizuno\altaffilmark{1,2}, Philip E. Hardee\altaffilmark{3}, and Ken-Ichi Nishikawa\altaffilmark{1,2}}

\altaffiltext{1}{Center for Space Plasma and Aeronomic Research, 
University of Alabama in Huntsville, 320 Sparkman Drive, 
NSSTC, Huntsville, AL 35805, USA; mizuno@cspar.uah.edu}
\altaffiltext{2}{National Space Science and Technology Center,
VP62, Huntsville, AL 35805, USA}
\altaffiltext{3}{Department of
Physics and Astronomy, The University of Alabama, Tuscaloosa, AL
35487, USA}

\shorttitle{3D RMHD simulations of CD instability with a sub-Alfv\'{e}nic jet}
\shortauthors{Mizuno et al.}

\begin{abstract}

We have investigated the influence of a velocity shear surface on the linear and non-linear development of the CD kink instability of force-free helical magnetic equilibria in 3D.  In this study we follow the temporal development within a periodic computational box and concentrate on flows that are sub-Alfv\'{e}nic on the  cylindrical jet's axis.  Displacement of the initial force-free helical magnetic field leads to the growth of CD kink instability. We find that helically distorted density structure propagates along the jet with speed and flow structure dependent on the radius of the velocity shear surface relative to the characteristic radius of the helically twisted force-free magnetic field.  At small velocity shear surface radius the plasma flows through the kink with minimal kink propagation speed. The kink propagation speed increases as the velocity shear radius increases and the kink becomes more embedded in the plasma flow. A decreasing magnetic pitch profile and faster flow enhance the influence of velocity shear.  Simulations show continuous transverse growth in the nonlinear phase of the instability. The growth rate of the CD kink instability and the nonlinear behavior also depend on the velocity shear surface radius and flow speed, and the magnetic pitch radial profile. Larger velocity shear radius leads to slower linear growth, makes a later transition to the nonlinear stage, and with larger maximum amplitude than occur for a static plasma column. However, when the velocity shear radius is much greater than the characteristic radius of the helical magnetic field, linear and non-linear development can be similar to the development of a static plasma column.
\end{abstract}

\section{Introduction}

Relativistic jets occur in black hole binary star systems (microquasars) (e.g., Mirabel \& Rodor\'{i}guez 1999), occur in active galactic nuclei (AGN) (e.g., Urry \& Padovani 1995; Ferrari 1998; Meier et al. 2001), can be associated with neutron stars and pulsar wind nebulae, e.g., Crab nebula jet (Weisskopf et al. 2000), and are thought responsible for the gamma-ray bursts (GRBs) (e.g., Zhang \& M\'{e}sz\'{a}ros 2004; Piran 2005; M\'{e}sz\'{a}ros 2006).  It is thought that these jets are powered and collimated hydromagnetically (e.g., Blandford 2000). Such Poynting flux dominated outflows arise on magnetic field lines threading the horizon of a rotating black hole (Blandford \& Znajek 1977; Blandford \& Payne 1982), and form as magnetic flux is pumped by the inflowing gas or generated by dynamo action in a surrounding accretion disk (e.g., Kudoh \& Kaburaki 1996). The observed high degree of jet collimation is generally attributed to hoop stress from the toroidal magnetic field acting in concert with an external pressure.

Jet generation simulations using general relativistic magnetohydrodynamics (GRMHD) (e.g., De Villiers et al. 2003, 2005; Hawley \& Krolik 2006, McKinney \& Gammie 2004; McKinney 2006; McKinney \& Blandford 2009; Beckwith et al. 2008; Hardee et al. 2007; Komissarov \& Barkov 2009; Penna et al. 2010) codes show development of magneto-rotational instability (MRI) (Balbus \& Hawley 1998) and angular momentum transfer in the accretion disk, leading to diffusion of matter and magnetic field inwards, and unsteady outflows near a centrifugally supported funnel wall. In general, GRMHD simulations with spinning black holes indicate jet production consisting of a Poynting-flux high Lorentz factor spine with $v \sim c$, and a matter dominated sheath with $v \sim c/2$ possibly embedded in a lower speed, $v << c$, disk/coronal wind. Circumstantial evidence such as the requirement for large Lorentz factors suggested by the TeV BL Lacs when contrasted with much slower observed motions (Ghisellini et al. 2005) suggests such a spine-sheath morphology, although alternative interpretations are also possible (Georganopoulos \& Kazanas 2003; Stern \& Poutanen 2008; Bromberg \& Levinson 2009;  Giannios et al. 2009).

Strongly magnetized relativistic outflows are typically produced from rotating bodies (neutron stars, black holes and accretion disks). A toroidal magnetic field ($B_{\phi}$) is wound up in such outflows and in the far zone becomes dominant because the poloidal field ($B_{p}$) falls off faster with expansion and distance. In configurations with strong toroidal magnetic field, the current-driven (CD) kink mode is unstable. This instability excites large-scale helical motions that can strongly distort or even disrupt the system. For static cylindrical force-free equilibria, the well-known Kruskal-Shafranov criterion indicates that the instability develops if the length of the plasma column, $\ell$, is long enough for the field lines to go around the cylinder at least once (e.g., Bateman 1978): $|B_{p}/B_{\phi}| < \ell/ 2 \pi R$.  For relativistic force-free static configurations, the linear instability criteria have been studied by several authors (Istomin \& Pariev 1994, 1996; Begelman 1998; Lyubarskii 1999; Tomimatsu et al. 2001; Narayan et al. 2009). In a more realistic case, rotation and shear could significantly affect the instability criterion.

Twisted structures are observed in many AGN jets on sub-parsec, parsec and kiloparsec scales (e.g., G\'{o}mez et al.\ 2001; Lobanov \& Zensus 2001). Non-relativistic and relativistic simulations of magnetized jet formation and/or propagation have showed helical structures attributed to CD kink instability (e.g., Lery et al.\ 2000; Ouyed et al.\ 2003; Nakamura \& Meier 2004; Nakamura et al.\ 2007; Moll et al.\ 2008; Moll 2009; McKinney \& Blandford 2009; Carey \& Sovinec 2009). In the absence of CD kink instability and resistive relaxation, helical structures may be attributed to the Kelvin-Helmholtz (KH) helical instability driven by velocity shear at the boundary between the jet and the surrounding medium (e.g., Hardee 2004, 2007) or triggered by precession of the jet ejection axis (Begelman et al.\ 1980). It is still not clear whether current driven, velocity shear driven or jet precession is responsible for the observed structures, or whether these different processes are responsible for the observed twisted structures at different spatial scales.  

In a previous paper we performed 3D relativistic MHD simulations and investigated the development of CD kink instability in a static plasma column with a force-free helical magnetic field  (Mizuno et al. 2009a). We found that the initial configuration was strongly distorted but not disrupted by the kink instability. Although static configurations (or more generally rigidly moving flows considered in the proper reference frame) are the simplest ones for studying the basic properties of the kink instability, in a realistic case, rotation and shear motions could significantly affect the CD kink instability. In this paper we investigate the influence of a velocity shear surface on the stability and nonlinear behavior of relativistic sub-Alfv\'{e}nic flow.  Because sub-Alfv\'{e}nic flow is stable to the Kelvin-Helmholtz (KH) instability, we can focus on the development of CD kink instability. In this paper we describe the numerical method and setup used for our simulations in \S 2, present our results in \S 3, in \S 4 compare our results to instability expectations and conclude, and in the Appendix present multi-dimensional numerical code tests.

\section{Numerical Method and Setup}

In order to study time evolution of the CD kink instability in the relativistic MHD (RMHD) regime, we use the 3D GRMHD code ``RAISHIN'' in Cartesian coordinates. RAISHIN is based on a $3+1$ formalism of the general relativistic conservation laws of particle number and energy-momentum, Maxwell's equations, and Ohm's law with no electrical resistance (ideal MHD condition) in a curved spacetime (Mizuno et al.\ 2006).  In the RAISHIN code, a conservative, high-resolution shock-capturing scheme is employed. The numerical fluxes are calculated using the HLL approximate Riemann solver, and flux-interpolated constrained transport (flux-CT) is used to maintain a divergence-free magnetic field
\footnote{Constrained transport schemes are used to maintain divergence-free magnetic fields in the RAISHIN code. This scheme requires the magnetic field to be defined at the cell interfaces. On the other hand, conservative, high-resolution shock capturing schemes (Godonov-type schemes) for conservation laws require the variables to be defined at the cell center. In order to combine variables defined at these different positions, the magnetic fields at the cell interfaces are interpolated to the cell center and as a result the scheme becomes
non-conservative even though we solve the conservation laws (Komissarov 1999).} . 
The RAISHIN code performs special relativistic calculations in Minkowski spacetime by choosing the appropriate metric. The RAISHIN code has proven to be accurate to second order and has passed a number of numerical tests including highly relativistic cases and highly magnetized cases in both special and general relativity (Mizuno et al.\ 2006). The results of multidimensional numerical test problems are shown in the Appendix. The multidimensional numerical test problems show that the MC slope-limiter scheme performs best on most of the tests and we have used this scheme for reconstruction.

In our simulations we choose a force-free helical magnetic field for the initial configuration (Mizuno et al. 2009a). 
A force-free configuration is a reasonable choice for the strong magnetic field cases that we study here. In general, the force-free equilibrium of a static cylinder is described by the equation
\begin{equation}
B_{z} {d B_{z} \over dR} + {B_{\phi} \over R} {d B_{\phi} \over dR} =0.
\end{equation}
In particular we choose the following form for poloidal ($B_{z}$) and toroidal ($B_{\phi}$) components of magnetic field determined in the laboratory frame
\begin{equation}
B_{z}= {B_{0} \over [1+ (R/a)^{2}]^{\alpha}}~,
\end{equation}
\begin{equation}
B_{\phi}= {B_{0} \over (R/a)[1+ (R/a)^{2}]^{\alpha}} \sqrt{ { [1 +
(R/a)^{2}]^{2 \alpha} -1 - 2 \alpha (R/a)^2 \over 2 \alpha -1}}~,
\end{equation}
where $R$ is the radial position in cylindrical coordinates normalized by a simulation scale unit $L \equiv 1$, $B_{0}$ parameterizes the magnetic field amplitude, $a$ is the characteristic radius of the column, and
$\alpha$ is the pitch profile parameter. The pitch parameter, $P \equiv R B_{z} / B_{\phi}$, determines the radial profile of the magnetic pitch, and larger $P^{-1}$ indicates increasing helical pitch of the magnetic field lines. With our choice for the force-free field, the pitch parameter can be written as
\begin{equation}
P = (R/a)^{2} \sqrt{ { 2 \alpha -1 \over [1 + (R/a)^{2} ]^{2 \alpha}
-1 -2 \alpha (R/a)^{2}} }~.
\end{equation}
If the pitch profile parameter $0.5 < \alpha < 1$, the magnetic helicity increases with radius. When $\alpha =1$, the magnetic helicity is constant and if $\alpha > 1$, the magnetic helicity decreases. This definition for the pitch parameter and the force-free helical field has been chosen to be the same as that used in previous non-relativistic work (Appl et al.\ 2000; Baty 2005) and in our previous relativistic work (Mizuno et al.\ 2009a) for purposes of comparison. For our modestly relativistic sub-Alfv\'enic flow speeds, the jet density and magnetic fields within the jet when determined in the jet flow frame are reduced slightly relative to their values in the laboratory frame by the flow Lorentz factor (e.g., Anile 1989; Komissarov 1999).

The simulation grid is periodic along the axial z direction. As a consequence the allowed axial wavelengths are restricted to $\lambda = L_{z}/n \le L_{z}$, with $n$ a positive integer and $L_{z}$ is the grid length. The grid is a Cartesian ($x, y, z$) box of size $4L \times 4L \times 3L$ with grid resolution of $\Delta L = L/40$. 
The grid resolution is the same in all directions. In simulations we choose a characteristic radius, $a=(1/4)L$. 
In terms of $a$, the simulation box size is $16a \times 16a \times 12a$ and the allowed axial wavelengths are restricted to $\lambda = 12a/n \le 12a$. We impose outflow boundary conditions on the transverse
boundaries at $x = y = \pm 2L$ ($\pm 8a$). This simulation grid is the same as that used for case B in Mizuno et al. (2009a). We checked the influence of  grid resolution for the case of a static plasma column using four different grid resolutions from 20 to 60 computational zones per simulation length unit $L=8a$ in Mizuno et al. (2009a). We found that  growth does depend on grid resolution, and our choice of $\Delta L=L/40$ is sufficient to capture the growth of CD kink instability.

We consider a low gas pressure medium with constant $p=p_{0} =0.02$ in units of $\rho_{0}c^{2}$ for the equilibrium state, and a non-uniform density profile decreasing with the magnetic field strength as, $\rho = \rho_{1} B^2$ with $\rho_{1} =6.25 \rho_{0}$. We choose a density decreasing $\propto B^{2}$ in order to keep the Alfv\'{e}n speed inside the velocity shear surface above the flow speed. The equation of state is that of an ideal gas with $p=(\Gamma -1) \rho e$, where $e$ is the specific internal energy density and the adiabatic index $\Gamma=5/3$
\footnote{
This adiabatic index assumes the plasma is cold, $c_{s} \ll c$. This condition is generally fulfilled in our simulations even though the sound speed is greater than the relativistic limit of $c/ \sqrt{3}$ at large radial position (see Fig. 1f). The region where the CD kink instability occurs is near the axis and in this region an adiabatic index $\Gamma=5/3$ is appropriate. We have checked the dependence of our simulation results on different adiabatic indicies (see Appendix A in Mizuno et al. 2009a). We found no significant difference in results for different adiabatic indicies.}. 
The specific enthalpy is $h \equiv 1+e/c^{2} +p/\rho c^{2}$. The magnetic field amplitude is $B_{0} =0.4$ in units of  $\sqrt{4\pi\rho_{0}c^{2}}$ leading to a low plasma-$\beta$ near the axis. The sound speed is $c_{s}/c \equiv (\Gamma p/\rho h)^{1/2}$ and the Alfv\'{e}n speed is given by $v_{A}/c  \equiv [B^{2}/(\rho h +B^{2})]^{1/2}$.

We choose two different jet velocities: (case s) slow, $v_{j}=0.2c$ and  (case f) fast, $v_{j}=0.3c$.  
In order to study the effect of the velocity, we perform the simulations with four different velocity shear surface radii: $R_{j}=1/8 L$ ($a/2$) , $1/4 L$ ($a$), $1/2 L$ ($2a$), $1 L$ ($4a$).  Results are compared to those for a static plasma column (no flow) as the reference. We also investigate the effect of different magnetic pitch profiles with: (case CP) constant pitch, $\alpha=1$, and (case DP) decreasing pitch, $\alpha=2.0$. The radial profiles of the magnetic field components, the magnetic pitch, and the sound and Alfv\'{e}n speeds determined in the lab frame for the different cases are shown in Figure 1. For all cases the sound and Alfv\'en speeds on the axis are $c_{s0} = 0.178~c$ and  $v_{A0} = 0.364~c$, respectively, when determined in the lab frame.
\begin{figure}[h!]
\epsscale{0.75}
\plotone{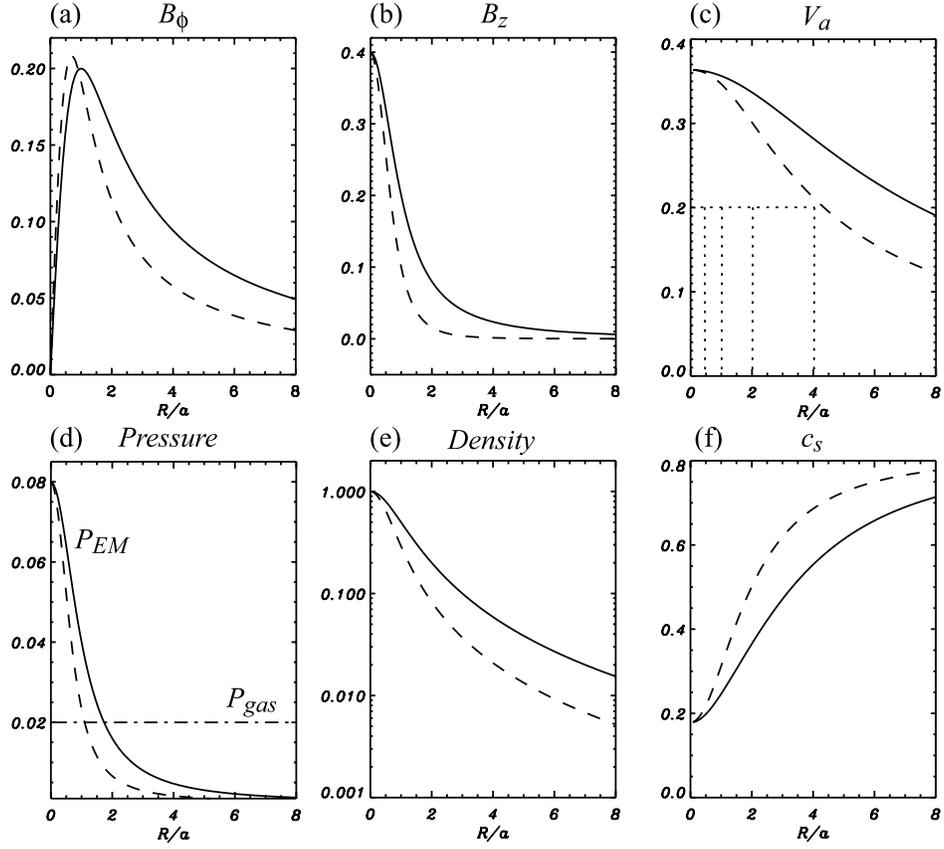}
\caption{Radial profile of (a) the toroidal magnetic field $B_{\phi}$, (b) the axial magnetic field $B_{z}$, (c) the Alfv\'{e}n speed, $v_A/c$, (d) magnetic and gas (dash-dot line)
pressure, (e) the rest mass density, $\rho$, and (f) the sound speed, $c_s/c$, all determined in the lab frame. Solid lines indicate the constant pitch case and dashed lines indicate the decreasing pitch case. Dotted lines in (c) show the locations of the velocity shear radius, $R_{j}/a=0.5$, $1.0$, $2.0$, and $4.0$ at $v_{j} = 0.2c$.
\label{f1}}
\end{figure}
We have chosen constant and decreasing pitch cases only because previous simulation studies of the CD kink instability of a static plasma column show that increasing magnetic pitch is not too different from the constant pitch case (Mizuno et al.\ 2009a). The jet velocity is sub-Alfv\'{e}nic inside the velocity shear radius for almost all cases with the exception of  the constant pitch case with the largest shear radius $R_{j}=4a$ and fastest jet velocity $v_{j}=0.3c$. In this paper we focus on the development of CD kink instability as the sub-Alfv\'{e}nic jet is stable to the velocity shear driven Kelvin-Helmholtz (KH) instability.

To break the symmetry the initial MHD equilibrium configuration is perturbed by a radial velocity in all region with profile given by
\begin{equation}
v_{R} = \delta v \exp \left(-  { R \over R_{p} } \right) \cos (m
\theta) \sin \left( {2 \pi n z \over L_{z}}\right)~.
\end{equation}
The amplitude of the perturbation is $\delta v = 0.01c$ with radial width $R_{p}= 0.5L$ ($2a$), and we choose $m=1$ and $n=1$.  This is identical to imposing $(m,n)=(- 1, -1)$, because of the symmetry between $(m,n)$
and $(-m, -n)$ pairs. The various different cases that we have considered are listed in Table 1.

\begin{deluxetable}{lcccc}
\tablecolumns{7}
\tablewidth{0pc}
\tablecaption{Models and Parameters}
\label{table1}
\tablehead{
\colhead{Case} & \colhead{$\alpha$} & \colhead{$v_{j}/c$} &
\colhead{$R_{j}/a$} & \colhead{Pitch}
} 
\startdata
CPsa/2 & 1.0 & 0.2 & 0.5 & constant \\
CPsa & 1.0 & 0.2 & 1.0  & constant \\   
CPs2a & 1.0 & 0.2 & 2.0   & constant \\
CPs4a & 1.0 & 0.2 & 4.0   & constant \\
CPfa/2 & 1.0 & 0.3 & 0.5 & constant \\
CPfa & 1.0 & 0.3 & 1.0  & constant \\   
CPf2a & 1.0 & 0.3 & 2.0   & constant \\
CPf4a & 1.0 & 0.3 & 4.0   & constant \\
CP0  & 1.0 & 0.0 & 0.0 & constant \\
DPsa/2 & 2.0 & 0.2 & 0.5 & decrease \\
DPsa & 2.0 & 0.2 & 1.0  & decrease \\   
DPs2a & 2.0 & 0.2 & 2.0   & decrease \\
DPs4a & 2.0 & 0.2 & 4.0   & decrease \\
DP0  & 2.0 & 0.0 & 0.0 & decrease \\
\enddata
\end{deluxetable}

\section{Results}

\subsection{Constant Helical Pitch: $v_j = 0.2~c$}

Figure 2 shows the time evolution of a density isosurface for constant helical pitch with $v_{j}=0.2c$ and $R_{j}=2a$ (CPs2a) where the time, $t$, is in units of $ t_c \equiv L/c = 4a/c$ (light travel time across the largest velocity shear surface radius, $R_j = 4a$, considered in this study). 
\begin{figure}[h!]
\epsscale{0.70}
\plotone{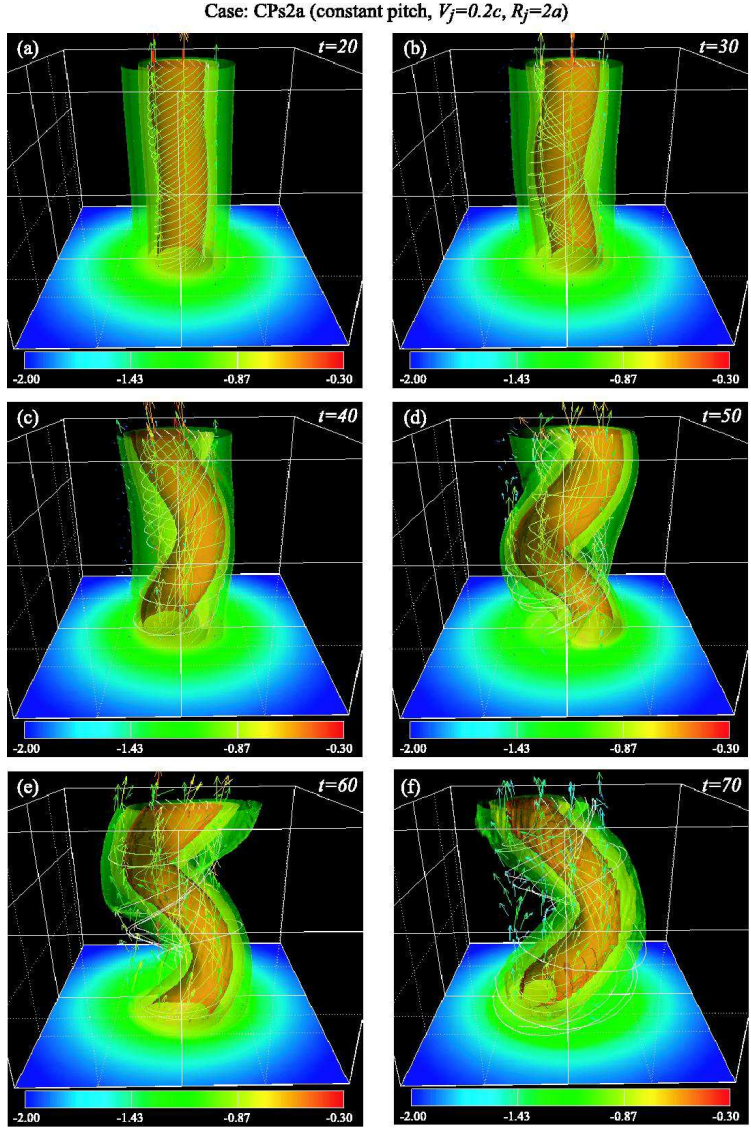}
\caption{Time evolution of three-dimensional density isosurfaces with a transverse slice at $z=0$ for case CPs2a. The time, $t$, is in units of $t_c = L/c$. Color shows the logarithm of the density with solid magnetic field lines. Velocity vectors are shown by the arrows. \label{A3m_3Dro}}
\end{figure}
Displacement of the initial force-free helical magnetic field by growth of the CD kink instability leads to a helically twisted magnetic filament wound around the density isosurface. In the nonlinear phase, helically distorted density structure shows continuous transverse growth and propagates in the flow direction. This propagation of the helical kink structure does not occur for a static plasma column (Mizuno et al. 2009a).

In order to investigate the dependence of kink growth and propagation on the location of the velocity shear surface, we consider four velocity shear radii from $R_{j}=a/2$ to $4a$ (see Table 1). The effect of different radii on the growth of the CD kink for constant magnetic pitch and $v_{j}=0.2c$ is shown in Figure 3. 
\begin{figure}[h!]
\epsscale{0.55}
\plotone{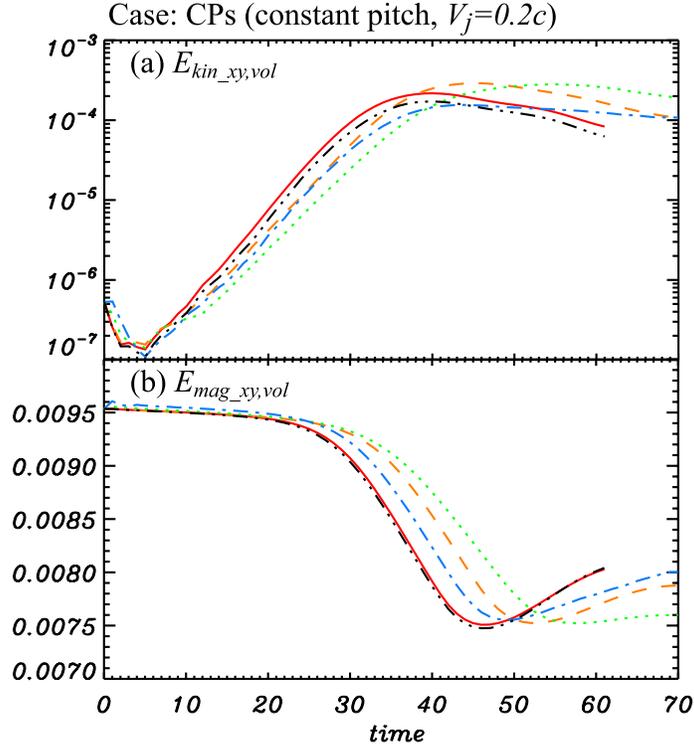}
\caption{Time evolution of (a) $E_{kin,xy}$ and (b) $E_{mag,xy}$ for constant pitch ($\alpha=1.0$) with $v_{j}=0.2c$ at different jet radii: $R_{j}/a=0.5$ (red solid line), $1.0$ (orange dashed line), $2.0$ (green dotted line) and $4.0$ (blue dash-dotted line). For reference a static plasma column case (no jet) is also shown as black dash double dotted lines. \label{tev_CPs}}
\end{figure}
In Figure 3 we show the time evolution of the volume-averaged kinetic, $E_{kin,xy}$, and magnetic, $E_{mag,xy}$, energy transverse to the $z$-axis determined within a cylinder of radius $R/L \le 1.0 ~(R \le 4a)$ as an indicator of the growth of the CD kink instability. 
The wavelength of the kink is $\lambda = 12a$. Note that according to the Kruskal-Shafranov criterion, the instability develops at $\lambda > 2\pi a$. The instability growth rate reaches a maximum at $\lambda_{max} \approx 10a$, the exact coefficient being dependent on the transverse distribution of the density and magnetic pitch, and also possibly the location and magnitude of the velocity shear.  For the case of constant pitch and uniform density Appl et al. (2000) found $\lambda_{max} = 8.43a$ and a corresponding growth rate of $\Gamma_{max} = 0.133 v_{A0}/a$. In general, one can use the estimate $\Gamma_{max} \approx 0.1 v_{A0}/a$ in the rest frame of the kink. For a moving kink we might expect the temporal growth rate in the lab frame to be reduced with $\Gamma \propto \gamma^{-1}_k$ where $\gamma_k$ is the moving kink Lorentz factor, e.g., Narayan et al. (2009).
Change in the evolution of $E_{kin,xy}$ and $E_{mag,xy}$ indicate an initial linear growth phase at  $t < (35 - 55) t_c$ with duration depending on the velocity shear radius, and followed by a nonlinear evolution phase. In all cases, the initial growth phase is characterized by an exponential increase in $E_{kin,xy}$ by about 3 orders of magnitude to a maximum amplitude followed by a slow decline in the nonlinear phase. 
By fitting the linear portion of the slope in $E_{kin,xy}$ between the amplitudes of $10^{-6}$ and $5 \times 10^{-5}$ we can determine an e-folding time where $\tau_e \equiv \Delta t/\ln 50$ and $\Delta t$ is the time interval. The e-folding times can be found in Table 2 in \S 4. In general, the e-folding time increases as the velocity shear radius increases for $a/2 < R_j \le 2a$. The e-folding time at $R_j =4a$ decreases but is still significantly longer than for the static plasma column. 
The time evolution trend of $E_{mag,xy}$ is opposite to the time evolution of $E_{kin,xy}$. $E_{mag,xy}$ gradually decreases in the early linear growth phase,  then exhibits an initial rapid decrease into the nonlinear phase to a minimum followed by a slight increase at later times.

At the smallest velocity shear radius the behavior of $E_{kin,xy}$ and $E_{mag,xy}$ are very similar to that of a static plasma column.  Increased difference in behavior from that of a static plasma column appears as the velocity shear radius increases to $R_{j}=a$ and $2a$.  For these cases as the velocity shear radius increases the growth rate of the CD kink slows, and $E_{kin,xy}$ achieves a somewhat higher maximum amplitude with a later transition to the nonlinear stage. As the shear radius increases $E_{mag,xy}$ exhibits a more gradual decline in the transition to the nonlinear stage than for the static plasma column. However, for the largest velocity shear radius, $R_{j}=4a$, the difference relative to the static plasma column in the linear and early non-linear phase is reduced, and the growth rate of the CD kink instability is faster and with lower maximum $E_{kin,xy}$ than for the cases with velocity shear radius $R_{j}=a$ and $R_{j}=2a$.  But note that $E_{kin,xy}$
becomes slightly larger than the static case at the longest comparable simulation times.

When the velocity shear radius is much larger than the characteristic radius of the helical magnetic field, we would expect the growth of CD kink instability to approach that of a static plasma column moving  with respect to the observer.  However, in the flow reference frame the Alfv\'en speed for the sub-Alfv\'{e}nic jet is not the same as that for a static plasma column in addition to a small relativistic clock effect. Thus, we do not expect the initial growth rate of largest jet radius case to perfectly match that of the static plasma column as determined in the observer (simulation) rest frame.  The effect of the velocity shear is greatest for the case with velocity shear radius $R_{j}=2a$ but also significant when $R_{j}=a$ and $4a$. 

Figure 4 shows a density isosurface at $t=50$ for the constant pitch cases with $v_{j}=0.2c$ at different velocity shear radii. In the linear growth phase, the behavior of the growing kink is almost the same for different velocity shear radius. However, in the nonlinear phase, the behavior of the kink is different for different velocity shear radius. For the smallest radius, $R_{j}=a/2$, the kink does not propagate along the z-axis significantly. Transverse amplitude growth dominates this case and is very similar to that of a static plasma column. In general the flow appears to follow the helical twist indicated by the density isosurface. Flow velocities in the transverse x-y plane are larger than flow along the z-axis.  When $R_{j}=a$ flow in the x-y plane is reduced somewhat relative to flow along the z-axis and the kink propagates slowly along the z-axis.
As the radius increases the kink propagates along the z-axis more rapidly (see more detail in Figure 5) while continuing transverse growth. Flow is less twisted helically as the velocity shear radius increases and at $R_{j} = 4a$ we see a helical kink embedded within and moving with the flow.  Recall that the time evolution of the volume averaged transverse kinetic energy, i.e., Fig. 3, were the most similar to a static plasma column for smallest and largest shear radius. The principal difference between the smallest and largest velocity shear radius cases lies in the flow morphology relative to the kink morphology indicated by the density isosurface. 
\begin{figure}[h!]
\epsscale{0.75}
\plotone{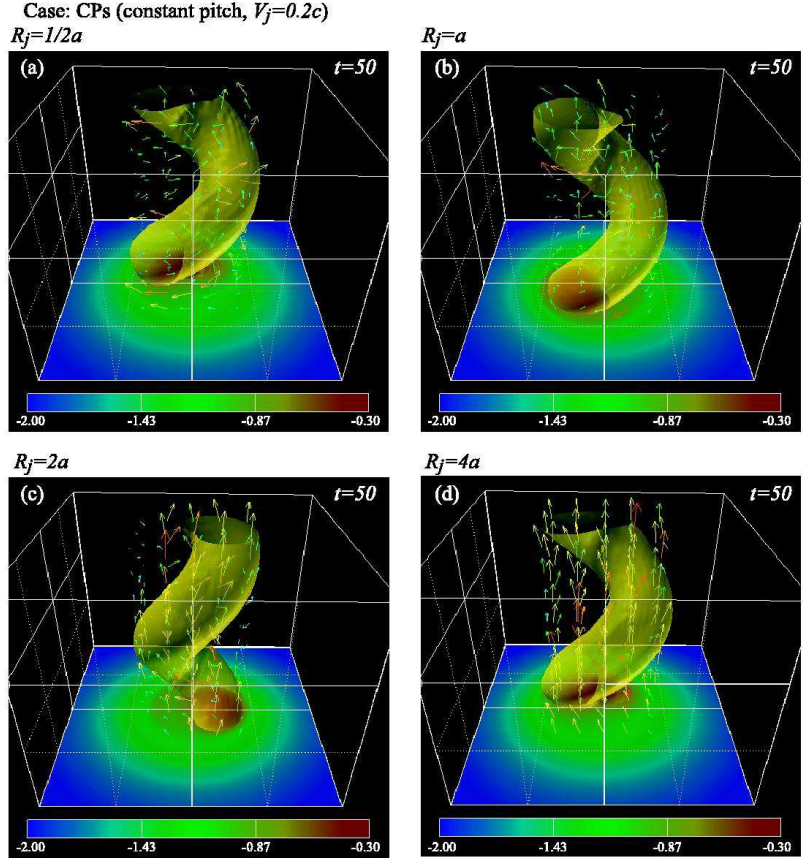}
\caption{Three-dimensional density isosurface at $t=50$ with a transverse slice at $z=0$ for constant helical pitch with $v_{j}=0.2c$ at different velocity shear radii: (a) $R_{j}/a=0.5$, (b) $1.0$, (c) $2.0$, and (d) $4.0$. Color shows the logarithm of the density with velocity vectors indicated by the arrows. \label{Am_3Dro}}
\end{figure}

In Figure 5 we follow the time evolution of the $x$ and $y-$position of the maximum density in the $xy$ plane at $z=1.5L = 6a$. Transverse growth of the kink is revealed by displacement of this maximum away from zero in the $xy$ plane. In all cases significant displacement of the density maximum begins at $t \gtrsim 20$, when growth is still within the linear regime. An oscillation of the maximum indicates rotation around the z-axis in the $xy$ plane and is related to the propagation speed of the kink.  Since the kink wavelength is $\lambda = 12 a$ in all simulations, the kink  propagation speed is $v_k = \lambda \nu = [12a/(\Delta t \times 4a)]c = (3/\Delta t)c$ where $\Delta t$ is the length of time for one rotation in simulation time units.
\begin{figure}[h!]
\epsscale{0.7}
\plotone{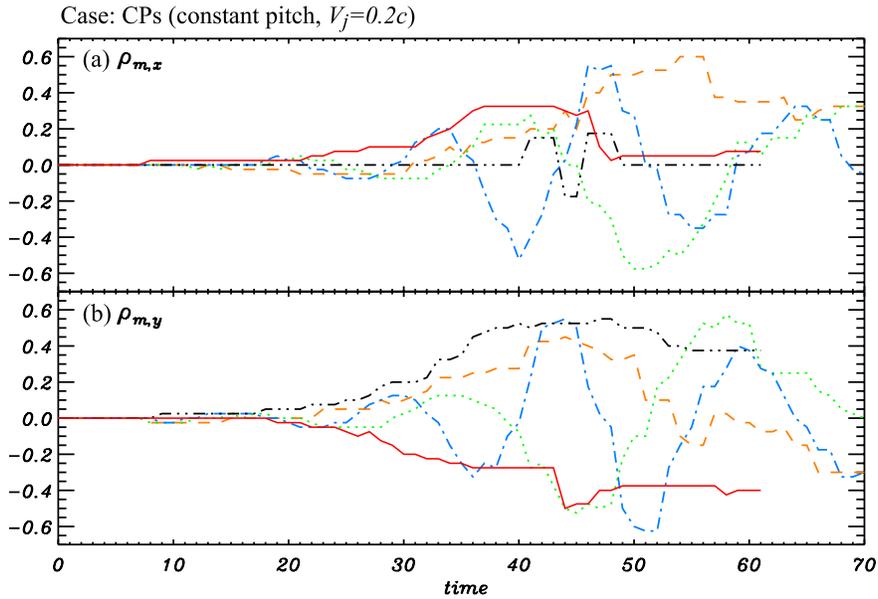}
\caption{Time evolution of (a) x and (b) y-position of maximum density in the $xy$ plane at $z/L=1.5$ ($6a$) for constant pitch ($\alpha=1.0$) with $v_{j}=0.2c$ at different shear radii: $R_{j}/a=0.5$ (red solid line), $1.0$ (orange dashed line), $2.0$ (green dotted line) and $4.0$ (blue dash-dotted line). The static plasma column is shown as a black dash-double dotted line. \label{tevp_CPs}}
\end{figure}

At the smallest shear radius (red solid lines), no oscillation is evident and transverse growth is very similar to that of the static plasma column (black dash double dotted lines). The lack of measurable propagation implies that the flow moves through the growing helical twist at nearly the flow speed, as suggested by the velocity vectors in panel (a) of Figure 4. When the shear radius is $R_{j}=a$ (orange dashed lines), only a partial oscillation occurs by the end of the simulation. Our best estimate of $\Delta t/2 \gtrsim 25$ comes from panel (b) in Figure 5 with a maximum and minimum displacement of $\rho_{m,y}$ occurring at $t \sim 44$ and $t \gtrsim 69$, respectively. This implies a kink propagation speed of $v_k/c \lesssim 0.06$. The kink is propagating along jet axis while transverse growth continues but the propagation speed is slow.  In this case the flow is still helically twisted but with less than the apparent helical twist of the density isosurface shown in panel (b) of Figure 4. 

More than one complete oscillation is evident for a shear radius $R_{j}=2a$ (see the green dotted lines) and visual inspection suggests that the oscillation period becomes longer with time. At $t < 45$ three measurements indicate $\langle \Delta t /2 \rangle \sim 10 \pm 1$, where ``$\pm$" indicates the range of the measurements. This implies a kink propagation speed $v_k /c \sim 0.15 \pm 0.01$. At $t > 40$ three measurements indicate $\langle \Delta t /2\rangle \sim 15 \pm 3$. This implies a kink propagation speed $v_k/c \sim 0.10 \pm 0.02$. The kink is propagating rapidly along the axis but as transverse growth continues the kink slows. In this case the flow is only modestly helically twisted and with much less than the apparent helical twist of the density isosurface shown in panel (c) of Figure 4. 

Multiple oscillations are evident for a shear radius 
$R_{j}=4a$ (see blue dash-dotted lines) and again the oscillation period becomes longer with time. Five measurements at  $t < 45$  provide $\langle \Delta t /2\rangle \sim 7.5 \pm 0.5$, and five measurements at $t > 45$  provide $\langle \Delta t/2\rangle \sim 8.5 \pm 1$. Here the initial kink speed  $v_k/c \sim 0.20 \pm 0.01$ is approximately equal to the flow speed. At later times the kink speed $v_k/c \sim 0.16 \pm 0.02$ has slowed significantly. At early time the smaller transverse amplitude of the kink is embedded within the flow and the flow exhibits little helical twist. The oscillation amplitude, indicative of  the transverse amplitude of the kink, increases up to about $t \sim 45$ and then declines slightly with time.  This suggests a decrease in transverse amplitude growth as the growing kink encounters the velocity shear surface and slows. 

We see from these results that kink propagation and flow morphology are strongly dependent on the velocity shear radius relative to the characteristic radius of the force-free plasma column.  At one extreme the fluid flows through the kink with helicity comparable to that of the kink and at the other extreme the kink is embedded within a more uniform flow. The kink propagation speeds are listed in Table 2 in \S 4.

\subsection{Decreasing Helical Pitch: $v_j = 0.2~c$}

In Figure 6 we show the time evolution of the volume-averaged kinetic, $E_{kin,xy}$, and magnetic, $E_{mag,xy}$, energy transverse to the $z$-axis determined within a cylinder of radius $R/L \le 1.0 ~(R \le 4a)$, e.g., Figure 3.
\begin{figure}[h!]
\epsscale{0.55}
\plotone{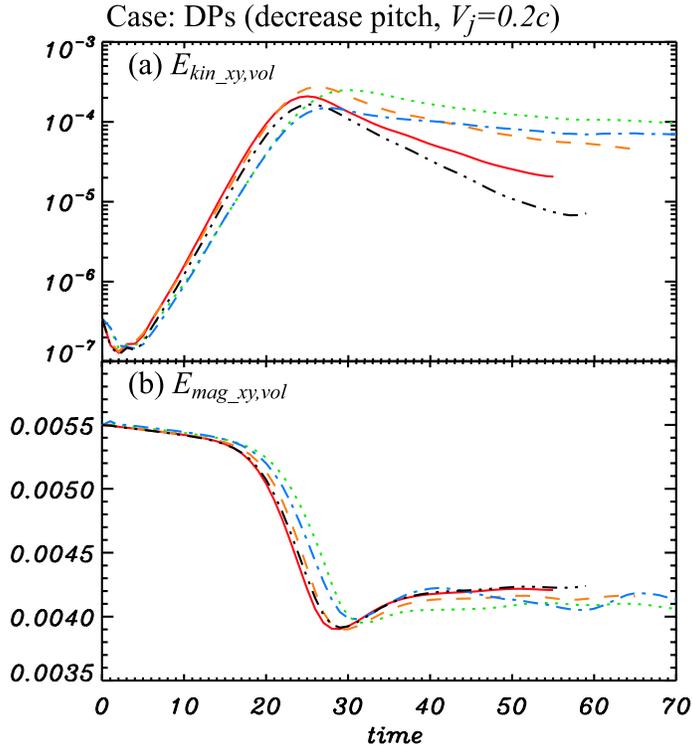}
\caption{Time evolution of (a) $E_{kin,xy}$ and (b) $E_{mag,xy}$ for decreasing helical pitch ($\alpha=2.0$) with $v_{j}=0.2c$ at different shear radii: $R_{j}/a=0.5$ (red solid line), $1.0$ (orange dashed line), $2.0$ (green dotted line) and $4.0$ (blue dash-dotted line). The static plasma column case (no flow) is also shown as black dash-double dotted lines. \label{tev_DPs}}
\end{figure}
Here we see an initial linear growth phase, $t < (25 - 30)t_c$, and subsequent nonlinear evolution phase. The  exponential increase in $E_{kin,xy}$ by about 3 orders of magnitude to a maximum amplitude followed by a slow decline in the nonlinear phase and behavior of $E_{mag,xy}$ is similar to what was found for the constant pitch cases. By fitting the linear portion of the slope in $E_{kin,xy}$ between the amplitudes of $10^{-6}$ and $5 \times 10^{-5}$ we can determine an e-folding time where $\tau_e \equiv \Delta t/\ln 50$ and $\Delta t$ is the time interval. The e-folding times can be found in Table 2 in \S 4. In this case, the e-folding time is increased relative to the static plasma column by about the same amount only for velocity shear radii  $2a \le R_j  \le 4a$. However, the decreasing helical pitch results in more rapid growth in the linear phase, and makes a transition to the nonlinear phase in about 60\% of the time for the constant pitch cases shown in Figure 3. This more rapid growth is similar to results for a static plasma column (Mizuno et al. 2009a). Thus, the growth rate trends found for the static plasma column are maintained in the presence of sub-Alfv\'{e}nic flow.  The maximum amplitude of $E_{kin,xy}$ is comparable to that found for constant helical pitch. However, the amplitude of $E_{kin,xy}$ declines less in the non-linear phase relative to the static case than was found for the constant pitch cases.  We note however that the constant pitch cases might have exhibited similar behavior at longer timescales. Here the decline in $E_{kin,xy}$ is clearly the least when $R_{j}=2a$ with decline increasing for $R_{j}= 4a$, $a$ and $a/2$, respectively.  The same trend is evident in Figure 3 for constant helical pitch. 

In this set of simulations we follow the development of the kink to a much longer time relative to the transition time from the linear growth phase to the non-linear phase.  Again we find that the largest effects occur for velocity shear radius
$R_{j}=2a$ but now confirm relatively large effects for $R_{j}=a$ and $R_{j} = 4a$. However, even for $R_{j}=a/2$ we clearly see the influence of velocity shear when compared with the results for constant pitch. We conclude that non-linear development is more influenced by a velocity shear surface in the case of decreasing helical pitch.

Figure 7 shows a density isosurface for the decreasing pitch cases with $v_{j}=0.2c$ for different velocity shear radius at $t=50$. 
\begin{figure}[h!]
\epsscale{0.75}
\plotone{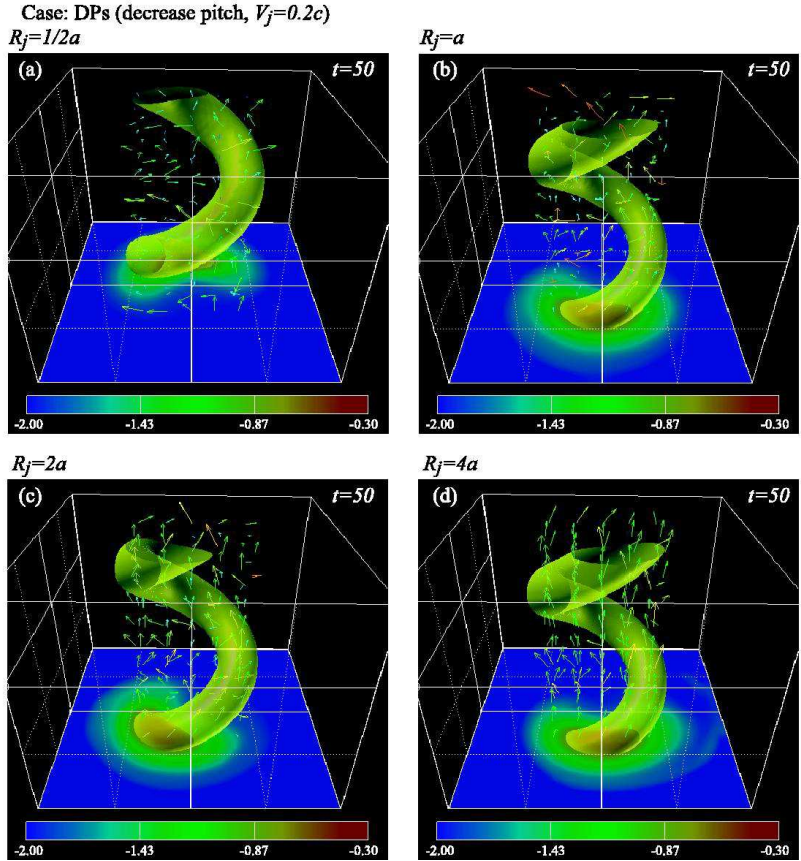}
\caption{Three-dimensional density isosurface at $t=50$ with transverse slices at $z=0$ for decrease pitch with $v_{j}=0.2c$ of different jet radius (a) $R_{j}/a=0.5$, (b) $1.0$, (c) $2.0$, and (d) $4.0$. Color shows the logarithm of the density with velocity vectors indicated by the arrows. \label{DPs_3Dro}}
\end{figure}
The decreasing pitch cases shown here all appear similar to results shown for static plasma columns (Mizuno et al. 2009a). In the linear growth phase the properties are almost same for different shear radius but after transition to the nonlinear phase the behavior of the kink is different for different shear radius. In general, the behavior is similar to what was found for constant helical pitch.  
For the smallest radius, $R_{j}=a/2$, the kink does not propagate along the z-axis significantly. In general the flow appears to follow the helical twist indicated by the density isosurface. Flow velocities in the transverse x-y plane are larger than flow along the z-axis.  When $R_{j}=a$ flow in the x-y plane is reduced somewhat relative to flow along the z-axis and the kink propagates slowly along the z-axis.
As the radius increases the kink propagates along the z-axis more rapidly (see more detail in Figure 8) while continuing transverse growth. Flow is less twisted helically as the velocity shear radius increases. However, we note that when $R_{j} = 4a$ we see considerably more indication of flow helicity than for the constant pitch case. 

In Figure 8 we follow the time evolution of the $x$ and $y-$position of the maximum density in the $xy$ plane at $z=1.5L = 6a$, e.g., Figure 5 for constant pitch cases. In all the decreasing pitch cases significant displacement of the density maximum 
begins at $t \lesssim 15$, when growth is still within the linear regime, and significant displacement occurs earlier than for the constant pitch cases.
\begin{figure}[h!]
\epsscale{0.8}
\plotone{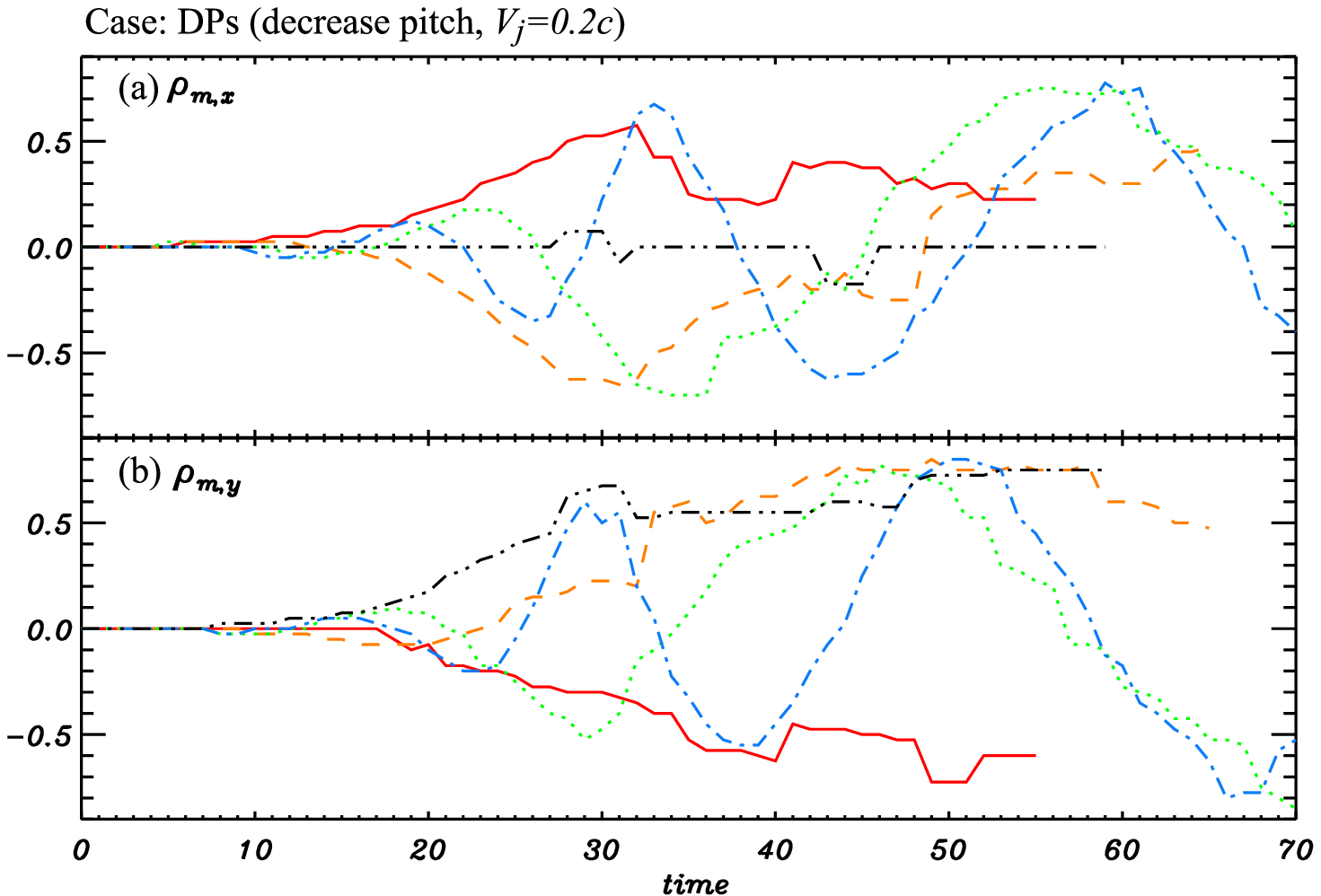}
\caption{Time evolution of (a) x and (b) y-position of maximum density in the $xy$ plane at $z/L=1.5$ ($6a$) for decreasing pitch ($\alpha=2.0$) with $v_{j}=0.2c$ oat different shear radii: $R_{j}/a=0.5$ (red solid line), $1.0$ (orange dashed line), $2.0$ (green dotted line) and $4.0$ (blue dash-dotted line). The static plasma column case (no flow) is shown as black dash-double dotted lines. \label{tevp_DPs}}
\end{figure}

At the smallest shear radius (red solid lines), no oscillation is evident and transverse growth is very similar to that of the static plasma column (black dash-double dotted lines). The lack of measurable propagation implies that the flow moves through the growing helical twist at nearly the flow speed, as suggested by the velocity vectors in panel (a) of Figure 7.
When the shear radius is $R_{j}=a$ (orange dashed lines), only a partial oscillation occurs by the end of the simulation. Our best estimate of $\Delta t/4 \sim 20$ comes from panel (a) in Figure 7 with a minimum and zero displacement of $\rho_{m,x}$ occurring at $t \sim 30$ and $t \sim 50$, respectively. This implies a kink propagation speed of $v_k/c \sim 0.04$. Comparison to the comparable constant pitch case suggests a reduction in the propagation speed but our level of accuracy does not make this a firm conclusion. The kink is propagating along jet axis while transverse growth continues but the propagation speed is slow.  In this case the flow is still helically twisted but with less than the apparent helical twist of the density isosurface shown in panel (b) of Figure 7. 

More than one complete oscillation is evident for a shear radius $R_{j}=2a$ (see the green dotted lines) and visual inspection suggests that the oscillation period becomes longer with time. At $t < 35$ two measurements indicate $\langle \Delta t /2 \rangle \sim 12.5 \pm 0.5$, where ``$\pm$" indicates the range of the measurements. This implies a kink propagation speed $v_k /c \sim 0.12 \pm 0.005$. At $t > 35$ two measurements indicate $\langle \Delta t /2\rangle \sim 24 \pm 1$. This implies a kink propagation speed $v_k/c \sim 0.0625 \pm 0.0025$.  The kink is propagating along the axis but as transverse growth continues the kink slows. Here the kink speed both at early and late times is significantly less than that found for the comparable constant pitch case.  

Multiple oscillations are evident for a shear radius 
$R_{j}=4a$ (see blue dash-dotted lines) and again the oscillation period becomes longer with time. Four measurements at  $t < 35$  provide $\langle \Delta t /2\rangle \sim 7.5 \pm 0.5$, and three measurements at $t > 35$  provide $\langle \Delta t/2\rangle \sim 14.5 \pm 1$. Here the initial kink speed  $v_k/c \sim 0.20 \pm 0.01$ is approximately equal to the flow speed. At later times the kink speed $v_k/c \sim 0.105 \pm 0.01$ has slowed significantly. At early time the smaller transverse amplitude of the kink is embedded within the flow and the flow exhibits little helical twist. The oscillation amplitude, indicative of  the transverse amplitude of the kink, increases up to about $t \sim 50$. Here we see transverse amplitude growth continuing to longer times and achieving larger amplitude than the comparable constant pitch case. The kink speed at later times is significantly less than that found for the comparable constant pitch case. 

We see from these results that kink propagation and flow morphology are again strongly dependent on the velocity shear radius relative to the characteristic radius of the force-free plasma column. In general, the presence of a velocity shear surface influences the propagation speed more than was found for the constant pitch cases.  The slower kink propagation speeds found for the decreasing pitch cases are likely the result of the faster amplitude growth of the kink to a larger transverse amplitude. Thus, at the larger velocity shear radii the growing kink approaches the velocity shear surface more rapidly and/or more closely. The kink propagation speeds are listed in Table 2 in \S 4.

\subsection{Constant Helical Pitch: $v_j = 0.3~c$ }

Figure 9 shows the time evolution of the volume-averaged kinetic energy ($E_{kin,xy}$) and magnetic energy ($E_{mag,xy}$) transverse to the $z$-axis within a cylinder of radius $R/L \le 1.0$ for constant pitch with $v_{j}=0.3c$, similar to Figures 3 and 6. Change in the evolution of $E_{kin,xy}$ and $E_{mag,xy}$ indicate an initial linear growth phase at  $t < (35 - 60)t_c$ with duration of the linear growth phase depending on the velocity shear radius. In all cases, the initial growth phase is characterized by an exponential increase in $E_{kin,xy}$ by about 3 orders of magnitude to a maximum amplitude followed by a slow decline in the nonlinear phase. 
By fitting the linear portion of the slope in $E_{kin,xy}$ between the amplitudes of $10^{-6}$ and $5 \times 10^{-5}$ we can determine an e-folding time. The e-folding times are listed in Table 2 in \S 4. In general, the e-folding time increases as the velocity shear radius increases for $a/2 < R_j \le 2a$. The e-folding time at $R_j =4a$ decreases but is still significantly longer than for the static plasma column. Evolution is qualitatively similar but  quantitatively different from the constant pitch slower jet cases. 
In particular, the maximum value for $E_{kin,xy}$ is about a factor of two larger than found for the constant pitch cases with $v_{j}=0.2c$, and is likely the result of the initial flow kinetic energy being about a factor of two higher.
\begin{figure}[h!]
\epsscale{0.55}
\plotone{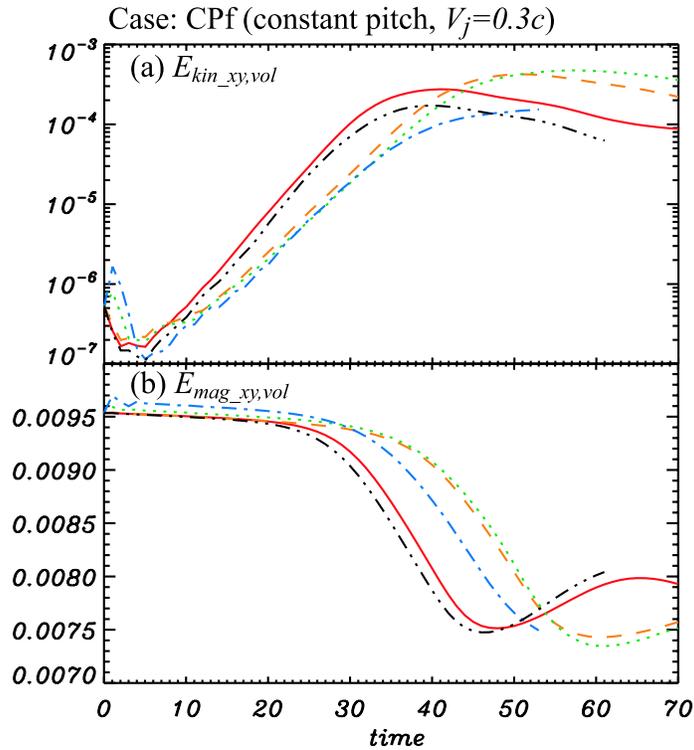}
\caption{Time evolution of (a) $E_{kin,xy}$ and (b) $E_{mag,xy}$ for constant pitch ($\alpha=1.0$) with $v_{j}=0.3c$ at different shear radii: $R_{j}/a=0.5$ (red solid line), $1.0$ (orange dashed line), $2.0$ (green dotted line) and $4.0$ (blue dash-dotted line). The static plasma column case (no flow) is shown as black dash-double dotted lines. \label{tev_CPf}}
\end{figure}
As was found for the slower jet cases, the growth rate is slower and transition to the nonlinear phase occurs later as the shear radius increases from $R_{j}=a/2$ (red lines) to $R_{j} = 2a$ (green dotted lines). As was found previously, the largest effects of velocity shear appear when $R_{j} = 2a$ with smaller but still significant effects when $R_{j} = a$. 

The $R_{j} = 4a$ simulation terminated before the maximum amplitude in $E_{kin,xy}$ was achieved, but it is clear that the maximum amplitude in $E_{kin,xy}$ will occur at significantly later time than when $R_{j} = a$. This result is different from the comparable slower flow case.  This difference is likely the result of relativistic effects and will be considered further in \S 4. 
The three-dimensional helical structure of these faster jet cases is qualitatively similar to that of slower jet cases with constant pitch. Density  isosurfaces, magnetic field lines and velocity vectors appear similar to those shown in Figures 2 \& 4. In Figure 10 we show  the time evolution of the $x$ and $y-$position of the maximum density in the $xy$ plane at $z=1.5L = 6a$, e.g., Figure 5 for the slower flow constant pitch cases. Here significant displacement of the density maximum begins at $t \sim 20$, when growth is still within the linear regime. This is similar to the slower flow constant pitch cases.
\begin{figure}[h!]
\epsscale{0.8}
\plotone{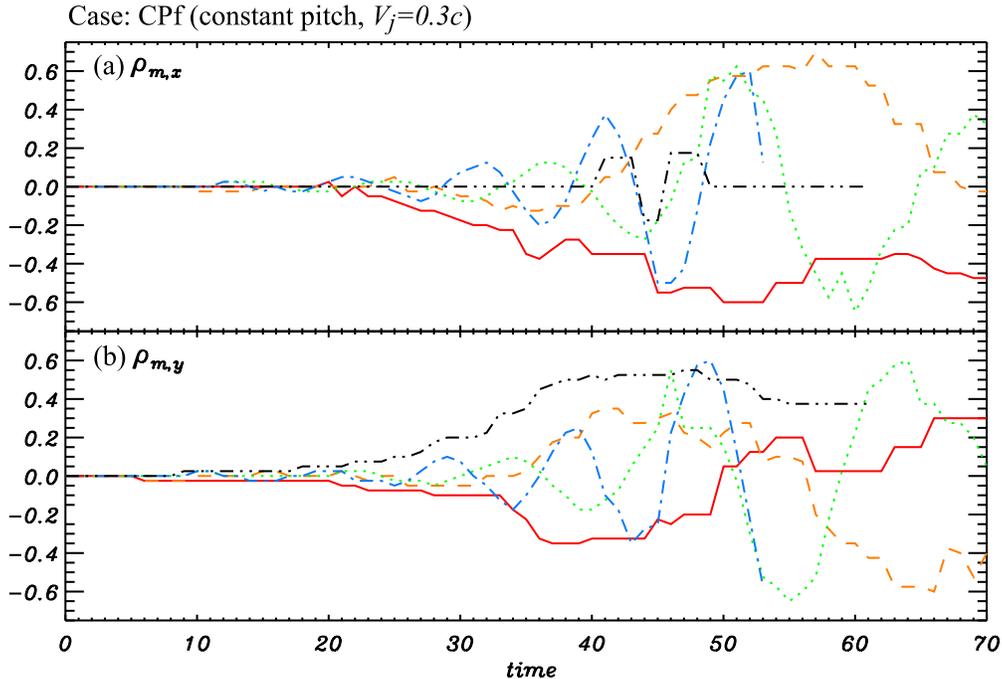}
\caption{Time evolution of (a) x and (b) y-position of maximum density in the $xy$ plane at $z/L=1.5$ ($6a$) for constant pitch ($\alpha=1.0$) with $v_{j}=0.3c$ at different jet radii: $R_{j}/a=0.5$ (red solid line), $1.0$ (orange dashed line), $2.0$ (green dotted line) and $4.0$(blue dash-dotted line). The static plasma column case (no flow) is also shown as black dash-double dotted lines. \label{tevp_CPf}}
\end{figure}
At the smallest shear radius (red solid lines), there is a suggestion of an oscillation in $\rho_{m,y}$ (panel b) with a maximum negative displacement at $t \sim 40$ and $\Delta t/2 \ge 30$, although no oscillation is evident in $\rho_{m,x}$ (panel a).  We can use this apparent oscillation to set an upper limit to the propagation speed of $v_k/c \le 0.05$.
Thus, the flow moves through the growing helical twist at nearly the flow speed, as was found for the previous cases. 

When the shear radius is $R_{j}=a$ (orange dashed lines), a full oscillation occurs and visual inspection suggests that the oscillation period increases with time. At $t < 35$ two measurements indicate $\langle \Delta t /2 \rangle \sim 12.75 \pm 1$, where ``$\pm$" indicates the range. This implies a kink propagation speed $v_k /c \sim 0.12 \pm 0.01$. At $t > 35$ two measurements indicate $\langle \Delta t /2\rangle \sim 24.75 \pm 2$. This implies a kink propagation speed $v_k/c \sim 0.06 \pm 0.005$.  The kink is propagating along the axis but as transverse growth continues the kink slows. Here the kink speed at late times is similar to that found for the comparable slower flow speed constant pitch case.  

Multiple oscillations are evident for a shear radius $R_{j}=2a$ (see the green dotted lines) and visual inspection suggests that the oscillation period becomes slightly longer with time. At $t < 45$ measurements indicate $\langle \Delta t /2 \rangle \sim 6.5 \pm 0.5$, where ``$\pm$" indicates the range. This implies a kink propagation speed $v_k /c \sim 0.23 \pm 0.015$. At $t > 45$ measurements indicate $\langle \Delta t /2\rangle \sim 8.75 \pm 0.5$. This implies a kink propagation speed $v_k/c \sim 0.17 \pm 0.01$.  The kink is propagating relatively rapidly along the axis at early times but as transverse growth continues the kink slows.  

Multiple oscillations are evident for a shear radius 
$R_{j}=4a$ (see blue dash-dotted lines) but there is no evidence that the oscillation period becomes longer with time for $t < 53$ when the simulation terminated. The multiple oscillations  provide $\langle \Delta t \rangle \sim 9.8 \pm 0.4$. Here the kink speed  $v_k/c \sim 0.305 \pm 0.01$ is approximately equal to the flow speed.  The oscillation amplitude, indicative of  the transverse amplitude of the kink, increases up to the end of this simulation. This suggests that the kink remains embedded in the flow up to the end of this simulation. 

These results are similar to those found for the slower flow cases with constant pitch.
When the shear radius is larger, the propagation speed of the helical kink becomes faster.  Quantitative comparison with individual slower jet cases (Fig. 5), on average indicates a more rapid propagation speed for each shear radius. The kink propagation speeds are listed in Table 2 in \S 4.

\section{Summary and Discussion}

We have investigated the development of the CD kink instability of a force-free helical magnetic field with a sub-Alfv\'{e}nic velocity shear surface located at various radii relative to the characteristic radius of the magnetic field. We restricted this study to sub-Alfv\'enic shear as this regime is appropriate to the magnetically dominated flows thought to exist in the acceleration and collimation regions of relativistic jets.  In this magnetically dominated parameter regime the flow is stable to the velocity shear driven Kelvin-Helmholtz instability so that we could focus solely on the effect of the shear flow on growth and propagation of the current driven kink and the velocity flow field accompanying the helically twisted kink.

The growth of CD kink instability in the initial exponential growth phase is slower than found for a static plasma column. In general, the reduction in the growth rate is larger for constant magnetic pitch than for decreasing magnetic pitch, and is larger if the velocity is larger.  In all cases, effects resulting from the presence of a velocity shear surface are largest when the velocity shear surface lies at or not too far outside, $R_j = a~\&~2a$,  the characteristic radius, $a$, of the force-free magnetic field. For the slower speed, $v_j = 0.2~c$, and constant pitch case it was clear that the initial growth was least affected when the velocity shear surface was inside, $R_j = a/2$, or far outside, $R_j = 4a$, the characteristic radius. For the higher speed, $v_j = 0.3~c$, and constant pitch case initial growth was again least affected when  $R_j = a/2$ but relativistic effects slowed the initial growth when $R_j = 4a$.
 
Transition to the nonlinear stage occurs at a later time but with larger maximum amplitude in the volume averaged transverse kinetic energy, $E_{kin,xy}$, as the velocity shear radius is increased for $R_j \le 2a$, when compared to a static plasma column. However, when the velocity shear radius is far from the characteristic radius, $R_{j} =  4a$, the maximum amplitude is comparable to that of the static plasma column although that maximum is reached after longer time.  In the absence of relativistic effects which slow the observed rate of growth, it is clear that the presence of a velocity shear surface has the strongest influence on both the linear and non-linear behavior of the growing kink when the velocity shear surface is located at or not too far outside the characteristic radius.    In general, decreasing the magnetic pitch or increasing the flow velocity enhances the influence of the velocity shear surface. 

The location of the  velocity shear surface has profound consequences for kink propagation and the associated flow field. For the velocity shear surface well inside  the characteristic radius, transverse growth is similar to the static plasma column.  In this case the plasma flows through the growing helical kink. For the velocity shear surface well outside the characteristic radius, the kink is embedded within and moves with the flow until the kink amplitude becomes large and the kink approaches the velocity shear surface.  In this case the initial transverse growth is similar to that of a static plasma column advected with the flow and with growth computed in the proper reference frame. As the growing  helically twisted kink approaches the velocity shear surface the kink slows and the flow field becomes slightly helically twisted.  For velocity shear radii on the order of the characteristic radius there is a more intimate interaction between the growing kink and the flow field.  In general, the kink propagates more slowly than the flow and slows as the amplitude increases.  Thus, the flow field becomes more helically twisted as the kink amplitude increases.  For these cases the flow helicity remains less than kink helicity. 

The Lorentz factors for slower and faster flows are $\gamma \simeq$~1.02 and 1.05, respectively. The growth rate of the CD kink instability depends on the Alfv\'{e}n velocity (e.g., Appl et al. 2000) and also depends on relativistic time dilation. In the flow reference frame, the toroidal and axial magnetic field components are reduced by the Lorentz factor and the mass density is reduced by the Lorentz factor squared. The Alfv\'{e}n velocity is  decreased slightly in the flow reference frame because $h = 1 + e/c^2 + P/\rho c^2$ is slightly larger. Therefore in faster flow cases, we would expect the growth rate to be reduced primarily by time dilation related to the Lorentz factor of the moving kink, $\gamma_k$. Our results for kink e-folding times in the linear growth phase are summarized in Table 2 along with the kink propagation speeds for the various different simulations.
\begin{deluxetable}{lcccc}
\tablecolumns{7}
\tablewidth{0pc}
\tablecaption{Models, e-folding times and kink speeds}
\label{table1}
\tablehead{
\colhead{Case} & \colhead{$v_{j}/c$} & \colhead{$R_{j}/a$} & \colhead{$\tau_e/t_{c}$\tablenotemark{a}} & \colhead{$v_k/c$\tablenotemark{b}}
} 
\startdata
CP0  & 0.0 & 0.0 & 3.75 & 0.0 \\
CPsa/2 & 0.2 & 0.5 & 3.75 & $\sim$ 0 \\
CPsa & 0.2 & 1.0 & 4.10  & 0.06 \\   
CPs2a & 0.2 & 2.0 & 4.35   & 0.15 - 0.10\tablenotemark{c} \\
CPs4a & 0.2 & 4.0 & 4.05   & 0.20 - 0.16\tablenotemark{c} \\
\tableline
DP0  & 0.0 & 0.0 & 2.45 & 0.0 \\
DPsa/2 & 0.2 & 0.5 & 2.40 & $\sim$ 0 \\
DPsa & 0.2 & 1.0 & 2.45  & 0.04 \\   
DPs2a & 0.2 & 2.0 & 2.75   & 0.12 - 0.06\tablenotemark{c} \\
DPs4a & 0.2 & 4.0 & 2.70   & 0.20 - 0.10\tablenotemark{c} \\
\tableline
CP0  & 0.0 & 0.0 & 3.75 & 0.0 \\
CPfa/2 & 0.3 & 0.5 & 3.70 & $<$ 0.05 \\
CPfa & 0.3 & 1.0 & 4.30  & 0.12 - 0.06\tablenotemark{c} \\   
CPf2a & 0.3 & 2.0 & 5.60   & 0.23 - 0.17\tablenotemark{c} \\
CPf4a & 0.3 & 4.0 & 5.30   & 0.30 \\
\enddata
\tablenotetext{a}{values indicated to nearest 0.05 and determined to $\pm$ 0.1}
\tablenotetext{b}{values determined to $\pm$ 5\%} 
\tablenotetext{c}{values at early - late simulation times} 
\end{deluxetable}

Quantitative comparison between our results and instability predictions is difficult because no sufficiently general stability analysis has been performed for magnetically dominated jets.  The stability analysis performed by Narayan et al. (2009) considers the case of a ``... rigid impenetrable wall at the outer cylindrical radius, $R_j$.'', that unfortunately is not appropriate to our simulations. Nevertheless, we can make some comparison with previous results for static plasma columns and consider the implications for the spatial development of the instability. In all cases, the initial axisymmetric structure is strongly distorted by the kink instability, even though not disrupted.  In general, the addition of a sub-Alfv\'enic velocity shear surface leads to slower temporal development of the instability than for the static case.  Comparison between the static and moving kink constant pitch cases shows that the e-folding times (see Table 2) are significantly longer, up to $\sim 16\%$ and $\sim 50\%$ longer for slow and fast velocity shear, when the shear surface is located at twice the characteristic radius of the plasma column.  When the velocity shear surface is located far outside the characteristic radius of the plasma column the kink is advected at about the flow speed in the linear growth regime and the e-folding times are shortened somewhat to about $\sim 8\%$ and $\sim 40\%$ longer than for the comparable static case.

With the kink moving with the flow frame we would expect the growth rate in the flow frame to be related to the growth rate in the lab frame by the Lorentz factor of the flow. However, comparison between our static plasma column kink e-folding times and our moving kink e-folding times is complicated by length contraction in addition to time dilation.  Time dilation increases the e-folding times from fluid to lab frames by the Lorentz factor. Length contraction means that our simulation box imposes a wavelength that appears longer in the fluid frame than in the lab frame by the Lorentz factor. In either frame the wavelength is longer than the fastest growing wavelength. Examination of our previous static case numerical results for the growth of $E_{kin_xy}$ at wavelengths of $\lambda = 12a$ and $16a$ (see Figure 2 in Mizuno et al. (2009a) for constant pitch cases A and B) indicates that the e-folding time increases about $7.5\%$ faster than proportional to the  wavelength over this wavelength range.  Thus, length contraction means that our e-folding time observed for a static kink should convert approximately to our e-folding time observed for a moving kink by $\tau^{mv}_e \sim 1.075 \gamma^2 \tau^{st}_e$. The e-folding times for cases CPs4a and CPf4a  when compared to the comparable static case, CP0, are significantly longer than the predicted 4\% ($v_k \sim 0.2~c$) and 11\% ($v_k \sim 0.3~c$) increase. We can only assume that a much larger velocity shear radius is required to further reduce the e-folding times for the propagating kinks to that predicted. On the other hand,  this also means that temporal kink growth is significantly slowed even for a velocity shear surface at four times the characteristic radius. It is interesting to note that the fluid inertia increases by $\gamma^2$ and if the growth time is also increased by the fluid inertia then the e-folding times in the lab frame should be increased by $\tau^{mv}_e \sim 1.075 \gamma^4\tau^{st}_e$ = 1.16 and 1.31 relative to the static case, and this increase comes much closer to the observed increase.  We speculate that the large increase in e-folding times measured in the lab frame is partly a result of the increased inertia of the relativistically moving fluid.

The characteristic time for the instability to affect strongly the initial structure varies from $(25-30) t_c$ for the decreasing pitch case to $(35 - 60) t_c$ for the constant pitch cases.  For the constant pitch cases the characteristic time is roughly $\tau \sim 10 \tau_e$, with values for $\tau_e$ being dependent on the structure of the undisturbed state. In a jet context our perturbations remain static or can propagate with the flow frame depending on the location of the velocity shear surface. In order to check whether the instability would affect a jet flow, one has to compare $\tau$ with a propagation time. 
If we identify $\tau_e$ with the fastest growing wavelength, our present results suggest a scaling like $\tau \sim 10 \gamma_k^{\alpha} \tau^{st}_e$ with $3 \ge \alpha \ge 1$ for a moving kink and with velocity shear surface a few times the characteristic radius. Here $\alpha=1$ would correspond to time dilation only and $\alpha = 3$ to time dilation plus inertial effects from the relativistically moving fluid.
In this case the condition for the instability to affect the jet structure might be written as
\begin{equation}
z  >  \gamma_k^{\alpha} v_k (A {a \over c})~,
\end{equation}
where we set $10 \tau^{st}_e \equiv A (a/ c)$  and $0 < v_k \le v_j$ is a function of $R_j/a$ and is sensitively dependent on the location of the velocity shear surface provided $R_j/a < < 10$. This result suggests that the characteristic scale for kink development could be longer or very much shorter than for a kink simply advected with a broad flow for which $z \gtrsim 100 \gamma_j  a$. 

In order to find a more general criteria one has to know how the characteristic radius, $a$, and Lorentz factor, $\gamma_k$, increase with distance.  If the jet is narrow enough so that $\Omega a^2/c < z$, where $\Omega$ is the angular velocity at the base of the jet, one can use the scaling (Tchekhovskoy et al. 2008; Komissarov et al. 2009; Lyubarsky 2009) $\gamma_j \sim \Omega a/c$ and assume that $\gamma_k = \epsilon \gamma_j$. In this case one finds that the criterion for the kink instability can be written as 
\begin{equation}
z c > A (\epsilon {\Omega a \over c})^{\alpha}[1 - (\epsilon {\Omega a \over c})^{-2}]^{1/2} a~.
\end{equation}
The instability could affect the jet structure only if the jet expands slowly enough and/or the kink moves slowly enough.  Assuming the parabolic shape for the jet, $\Omega a/c = \xi(\Omega z/c)^k \equiv \xi \chi^k$, where $k < 1$ and $\xi \sim 1$ are dimensionless numbers, one finds that the instability develops only if $k < 1/(\alpha + 1)$.  In this case the characteristic scale for the development of the instability can be written as
\begin{equation}
 \Omega z/c = \chi \sim (A \epsilon^{\alpha} \xi^{\alpha + 1})^{1/[1 - (\alpha + 1)k]}[1 - (\epsilon \xi \chi^k)^{-2}]^{1/2[1 - (\alpha +1)k]}~. 
\end{equation}
For $\epsilon = 1$,  $\alpha = 1$ and $v_k = v_j \sim c$ we recover the case for a kink advected with the flow field, eq.(9) in Mizuno et al. (2009a), $\Omega z/c \sim (10 \xi)^{2/(1 - 2k)}$ for A = 100. Here as $1 > \epsilon \rightarrow 1/\xi \chi^k$, $v_k/v_j < < 1$ and $\Omega z /c < 1$ so the characteristic scale for development of instability can be very short.  For $\alpha > 1$, i.e., potential inertial effects, the characteristic time is lengthened and this has the potential for lengthening the characteristic scale for the development of the instability.  This shows up in the constraint on $k < 1/(\alpha + 1) \le 1/2$, i.e., only if the jet expands slowly enough. 

The 3D relativistic jet generation simulation performed by McKinney \& Blandford (2009) indicates relatively rapid, less than 100 gravitational radii, but non-disruptive kink development over 500 gravitational radii. Our previous and present simulations for static and moving kinks suggest that the rapid but non-disruptive kink development in the jet generation simulation could be a result of a velocity shear surface located less or on order of the characteristic magnetic radius and a density increasing with radius.  This combination would result in a slowly moving kink, hence rapid initial spatial development, but with non-linear growth slowed by the density  increase and accompanying Alfv\'en speed decline with radius, increasing the Alfv\'en crossing time and slowing spatial development.  A non-linearly stabilizing increasing density profile is what might be expected for a Poynting flux jet core confined within a senser slowly moving sheath, as appears to be predicted by jet generation simulations. Of course, a proper investigation of spatial growth requires stability simulations designed to study spatial kink development using more realistic flow, magnetic, and density profiles.

In this paper, we considered sub-Alfv\'{e}nic jet flow and focused on the development of the CD kink instability. If the velocity shear is super-Alfv\'{e}nic, the flow can be KH unstable and CD unstable. Baty \& Keppens (2002) have investigated the interaction between KH and CD driven instabilities of a magnetized force-free cylindrical configuration via 3D MHD simulations in the non-relativistic regime. They found that the CD unstable modes provided a stabilizing effect on KH instability driven vortical structure. However, they assumed a relatively weak magnetic field in their simulations and studied the super-Alfv\'enic regime where the  KH instability grows faster than the CD kink instability (e.g., Appl et al. 2000; Baty 2005). If the magnetic field is strong but the velocity shear is weakly super-Alfv\'enic, the growth rate of the KH instability can be less than or comparable to that of the CD instability. Even in the super-Alfv\'enic regime the KH instability can be suppressed if a jet is embedded in a slower moving magnetized sheath that reduces the velocity shear to being effectively sub-Alfv\'enic (Hardee 2007; Mizuno et al. 2007).  Such a sheath may exist around jets in the acceleration and collimation region.  Thus, investigation of the weakly super-Alfv\'enic parameter regime will be particularly important to understanding the development of the twisted structures that are observed on relativistic jets. In future work we will investigate the spatial development of CD instability and the coupling between CD instability and KH instability.

\acknowledgments
Y.M. thanks Y. Lyubarsky, K. Shibata, B. Zhang, M. A. Aloy, and J. M. Stone for helpful discussions. 
This work is supported by NSF awards AST-0506719, AST-0506666, AST-0908010, and AST-0908040, and NASA awards NNG05GK73G, NNX07AJ88G, and NNX08AG83G, and US-Israeli BSF award 2006170. The simulations were performed on the Columbia Supercomputer at NAS Division at NASA Ames Research Center, the SGI Altix (cobalt) at the National Center for Supercomputing Applications in TeraGrid project which is supported by the NSF and the Altix3700 BX2 at YITP in Kyoto University.

\newpage

\appendix

\section{Multidimensional Numerical Tests}

In this section we first summarize previous code results against known solutions and second we check our numerical simulation code results against those obtained by other codes using the same test problems. In RMHD there are not many problems with known solutions because a closed solution for the general RMHD Riemann problem has not yet been found. 

The code has been verified against a number of test and physical problems with known solutions. In Mizuno et al.\ (2006), a 1D linear Alfv\'en wave propagation test and a 1D magnetized Bondi accretion flow test showed second-order convergence for our simulation code, and relativistic MHD shock-tube tests showed that our simulation code correctly handles the wave structure of both shocks and rarefactions even in an extreme relativistic regime.   The code has been used to model several 1D and 2D physical problems with known solutions.  The code was successfully used to solve the 1D Riemann problem for the deceleration of an arbitrarily magnetized relativistic flow injected into a static unmagnetized medium (Mizuno et al.\ 2009b). In particular, this physical problem involved comparing conditions for existence of the reverse shock against known relativistic shock jump conditions.  The code has also been successfully used to solve in 2D the effect of magnetic fields on an HD/MHD boost mechanism proposed by Aloy \& Rezzolla (2006).  The results to this physical problem  involving a relativistic flow that flows parallel to an overpressure generated shock and rarefaction wave combination agreed with and extended the previous HD findings to MHD (Mizuno et al.\ 2008).

We also have used our simulation code for several multidimensional (3D) physical problems involving comparison with known theoretical predictions. The code has been used to study spatial development of the Kelvin-Helmholtz instability on relativistic super-Alfv\'enic and trans-Alfv\'enic  cylindrical jets, and results (see Mizuno et al.\ 2007) were compared to predictions made by a linear stability analysis (Hardee 2007). The code has also been used to study temporal development of the current driven instability of a relativistic static plasma column, i.e., with magnetic energy density comparable to the plasma energy density including the rest mass energy, and results were successfully compared to linear stability analysis predictions and to previous non-relativistic numerical results (see Mizuno et al.\ 2009a).  In both completely different instability regimes, the code delivered results that agreed with stability analysis predictions and with previous numerical results. These successful comparisons verify the code against known solutions to physical multidimensional problems.

In general, we have not yet shown results for multidimensional test problems. In what follows we check our simulation code against multidimensional test problems using the MC slope-limiter scheme which we have used for the simulations in this paper. In all test simulations we use the HLL approximate Riemann solver to calculate numerical fluxes and the flux-CT scheme to maintain a divergence-free magnetic field.

\subsection{Advection of a Magnetic Field Loop}

We consider the advection of a weak magnetic field loop in an initially uniform velocity field. In thermal pressure dominance the loop is transported as a passive scalar. The 2D test that we perform is  a relativistic version of the magnetic loop advection problem. This tests the dissipative properties of the numerical scheme and the correct discretization balance of multidimensional terms (Gardiner \& Stone 2005, 2008; Mignone et al.\ 2010). A successful test results in the preservation of the initial loop.

Following non-relativistic MHD tests (Gardiner \& Stone 2005, 2008; Mignone et al.\ 2010), we employ a periodic computational box defined as $-0.3 \le x \le 0.3$ and $-0.15 \le y \le 0.15$ discretized on $N_{x} \times N_{x}/2$ computational zones. Density and gas pressure are initially constant, and set $\rho=1.0$ and $p=0.328$ respectively. The sound speed $c_{s} =0.3c$ for an adiabatic index $\Gamma=5/3$.
The magnetic field is defined through its magnetic vector potential as
\begin{equation}
A_{z} = \left\{
 \begin{array}{lll}
 a_{0}+a_{2}r^{2} & \mbox{if} & 0 \le r \le R_{1}, \\
 A_{0}(R-r) & \mbox{if} & R_{1} < r \le R, \\
 0 & \mbox{if} & r > R,
 \end{array}
\right.
\end{equation}
where $A_{0}=10^{-3}$, $R=0.09$, $R_{1}=0.2R$, $a_{2}=-0.5 A_{0}/R_{1}$, $a_{0}=A_{0}(R-R_{1})-a_{2}R_{1}^{2}$, and $r=\sqrt{x^{2}-y^{2}}$. This modification of the vector potential in the $r \le R_{1}$ region, with respect to the original version of the test problem performed by Gardiner \& Stone (2005), is done to remove the singularity in the loop's center that can cause spurious oscillations and erroneous evaluations of the magnetic energy (Mignone et al.\ 2010). We perform a case with no advection and an advection case in which the velocity of the flow is set up as $v_{x}=0.6c$, $v_{y}=0.3c$ and $v_{z}=0.0c$. The simulations are evolved until $t=1$ when the loop has crossed through the periodic boundaries and returned to the center of the grid in the advection case. 

In Figure 11 the magnetic energy density ($B^{2}$) and magnetic file lines are shown for the no advection case (upper) and advection case (lower) with $N_{x}=256$. In the no advection case, the circular shape of the loop is perfectly preserved. There is no visible distortion or dissipation around the center and outer boundaries of the field loop. In the advection case, the circular shape of the loop is relatively well preserved although some flattening of the magnetic field lines is seen at left upper and right lower outer boundaries to the field loop. The magnetic energy density image shows internal structure comparable to that found in the Newtonian limit by Gardiner \& Stone (2005) using first and second order CTU+CT schemes (Fig. 3 and 6 in Gardiner \& Stone 2005).  Mignone et al.\ (2010) performed a similar 2D magnetic loop advection test in the Newtonian limit using higher order schemes (third and fifth order). Not surprisingly, the higher order schemes do a better job of maintaining uniformity in the magnetic energy density and in the loop structure.
\begin{figure}[h!]
\epsscale{0.90}
\plotone{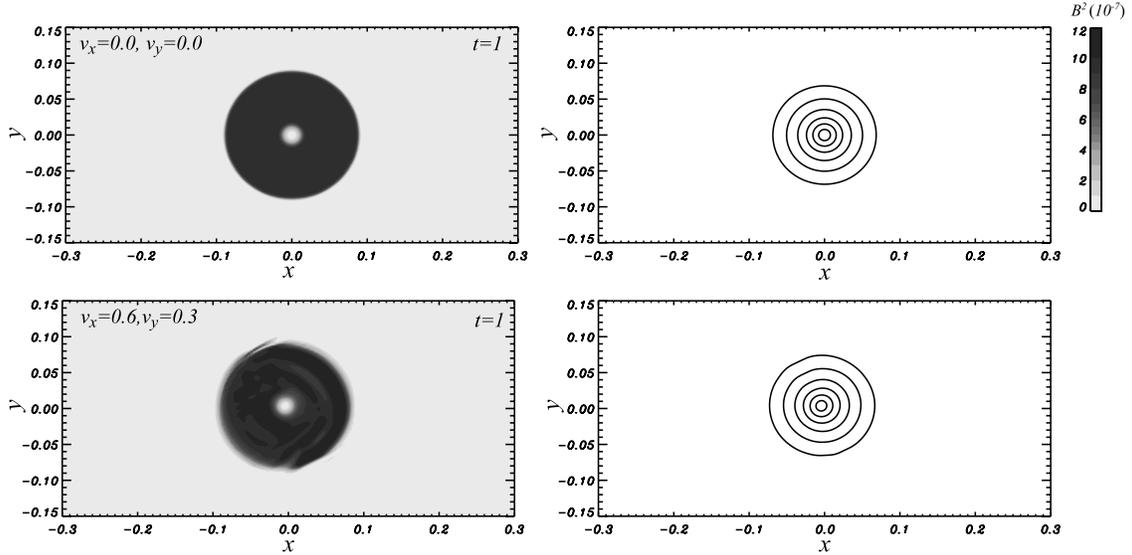}
\caption{Gray scale images of the magnetic energy density ($B^{2}$)({\it left}) and magnetic field lines ({\it right}) for the 2D magnetic field loop problem with no flow advection ({\it upper}) and with advection ({\it lower}) at $t=1$ for $N_{x}=256$.
 \label{fap1}}
\end{figure}

A quantitative measure of the magnetic field dissipation rate is indicated by the time evolution of the volume averaged magnetic energy density normalized to its initial value as shown in Figure 12. 
\begin{figure}[h!]
\epsscale{0.35}
\plotone{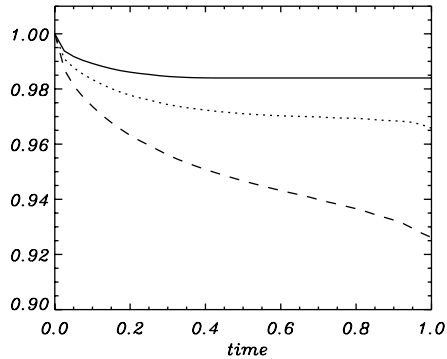}
\caption{Time evolution of the volume-averaged magnetic field energy density normalized to its initial value in the 2D field loop advection problem using the MC scheme with $N_{x}=512$ ({\it solid}), $256$ ({\it dotted}), and $128$ ({\it dashed}). 
 \label{fap2}}
\end{figure}
Higher numerical resolution leads to a less diffusive result with less than $2\%$ magnetic energy loss at the highest numerical resolution. Our result using $N_{x}=128$ finds almost the same magnetic energy dissipation, about 7\% magnetic energy loss, as that found by Gardiner \& Stone (2005) (see $t=1$ in Fig.\ 7 of  Gardiner \& Stone 2005) in the Newtonian limit using a comparable second-order scheme. Our quantitative results compare favorably with the higher order schemes tested by Mignone et al.\ (2010). Here our result using $N_{x}=128$ ($\sim 7\%$ loss) is similar to their result using a third-order scheme ($\sim 6.5\%$ loss) at the same resolution (see $t=1$ in Fig.A.9 of Mignone et al.\ 2010). Our result using higher resolution $N_{x}=256$ shows less magnetic energy dissipation ($\sim 3\%$ loss) at time $t=1$.

The three dimensional version of this problem is particularly challenging and see Gardiner \& Stone (2008) for comparable second-order scheme results in the Newtonian limit.  Correct evolution depends on how accurately the divergence-free condition is preserved and how well multidimensional MHD terms are balanced. The test consists of a computational box defined as $-0.15 \le x \le 0.15$, $-0.15 \le y \le 0.15$ and $-0.3 \le z \le 0.3$ discretized on $N_{x}/2 \times N_{x}/2 \times N_{x}$ computational zones with periodic boundary condition in all directions. Following the two-dimensional case, density and gas pressure are initially constant, and we set $\rho=1.0$ and $p=0.328$. We show a no advection case and an advection case in which the velocity of the flow is set as $(v_{x}, v_{y}, v_{z})=(0.3c, 0.3c, 0.6c)$. As for the two-dimensional case the vector potential  $A_{z}$ is used to initialize the magnetic field in the coordinate system $(x_1, x_2, x_3)$ which is related to computational coordinate system $(x, y, z)$ via the rotation
\begin{eqnarray}
\begin{array}{lll}
      x_1= \cos \gamma x - \sin \gamma z\\
      x_2= y\\
      x_3= \sin \gamma x + \cos \gamma z, 
\end{array}
\end{eqnarray}
where $\gamma=\tan^{-1} 1/2$. The initial vector potential is given by Eq. (A1), with $r=\sqrt{x_1^2 + x_2^2}$.

Figure 13 shows three-dimensional isovolume images of the magnetic energy density for the no advection (left) and advection case (right) with $N_{x}=256$. In the no advection case, the cylindrical shape of field loops is perfectly preserved. Some small dissipation is seen around the center and outer boundaries of the field loops. 
In the advection case, the cylindrical shape of the field loops is relatively well preserved although there is some distortion and dissipation around the inner and outer boundaries of the field loops.  This result is similar to that  seen in the 2D field loop advection case (see Fig.\ 11) and appears comparable to the results shown in Figure 2 in Gardner \& Stone (2008).   Mignone et al.\ (2010) performed a similar 3D magnetic loop advection test in the Newtonian limit using higher order schemes (third and fifth order). Not surprisingly, the higher order schemes do a better job preserving the loop structure and also display sharper boundaries.
\begin{figure}[h!]
\epsscale{0.80}
\plotone{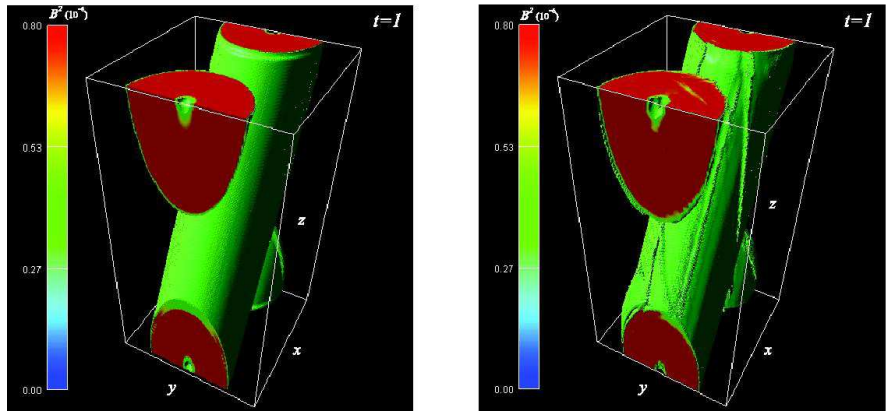}
\caption{Three-dimensional isovolume images of the magnetic energy density ($B^{2}$) for the 3D magnetic field loop problem ({\it a})  with no flow advection and ({\it b}) with advection at $t=1$ for $N_{x}=256$.
\label{fap2a}}
\end{figure}

A quantitative measure of the magnetic field dissipation rate in 3D field loop advection problem is indicated by the time evolution of the volume averaged magnetic energy density normalized to its initial value as shown in Figure 14. 
\begin{figure}[h!]
\epsscale{0.35}
\plotone{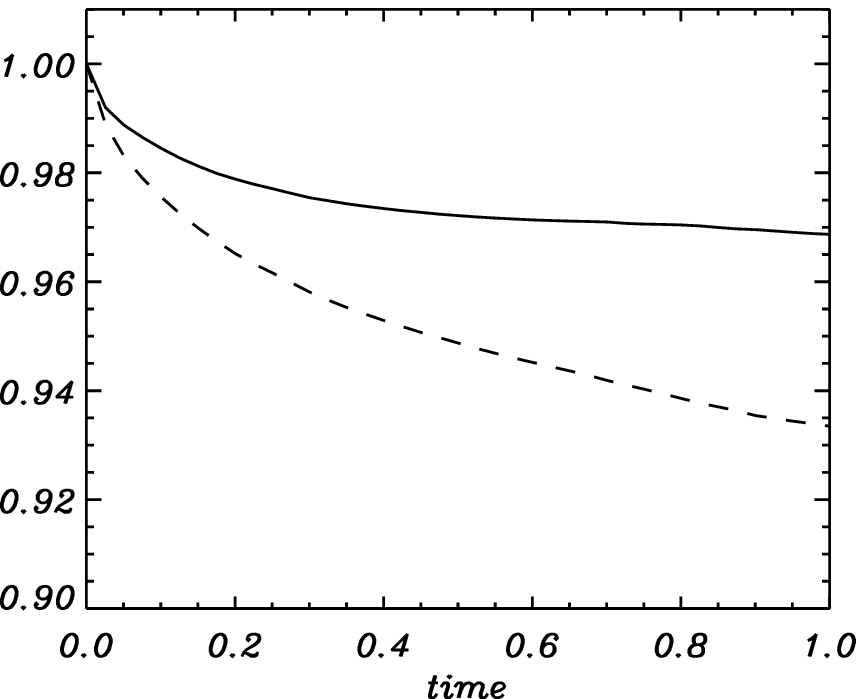}
\caption{Time evolution of the volume-averaged magnetic field energy density normalized to its initial value in the 3D field loop advection problem using the MC scheme with $N_{x}=256$ ({\it solid}) and $128$ ({\it dashed}). 
 \label{fap2b}}
\end{figure}
As found for the 2D magnetic advection problem, higher numerical resolution leads to a less diffusive result with $\sim 3\%$ magnetic energy loss at the highest numerical resolution of $N_{x}=256$. Our result using $N_{x}=128$ finds a slightly greater magnetic energy dissipation ($< 7\%$  loss) when compared to the $\sim 5\%$ magnetic energy loss found by Gardiner \& Stone (2008) (see Fig.\ 1 in  Gardiner \& Stone 2008) in the Newtonian limit using a comparable second-order scheme.  Our magnetic energy loss using $N_{x}=128$ ($< 7\%$ loss) is greater than that found by Mignone et al.\ (2010) (see Fig.A.9 in Mignone et al.\  2010) using a third-order scheme ($\gtrsim 3\%$ loss) at the same resolution. However, our result using higher resolution $N_{x}=256$ shows less magnetic energy dissipation ($\sim 3\%$ loss).

In summary, we obtain results very similar to previous non-relativistic results obtained using comparable second-order schemes (Gardiner \& Stone 2005; Gardiner \& Stone 2008). At present there are no published 2D or 3D relativistic loop test results for comparison. We conclude that our simulation code successfully passes both 2D and 3D test problems for the second-order MC slope-limiter and flux-CT schemes. We note that test simulations without the flux-CT scheme fail to pass these test problems.

\subsection{2D Cylindrical Explosion}

The cylindrical explosion test problem consists of a strong shock propagating into a magnetically dominated medium. Results for different cylindrical explosion problems in RMHD have been reported by several authors (e.g., Dubal 1991; van Putten 1995; Komissarov 1999; Del Zanna et al.\ 2003; Leismann et al.\ 2005; Mignone \& Bodo 2006; Ant\'{o}n et al.\ 2010). Here, unlike the 2D loop test we can compare our results to identical relativistic test results obtained by other RMHD second-order codes. This test verifies that the generation of a high Lorentz factor flow is handled correctly. A successful test results in a dramatic difference in pressure and density between the ambient gas and the explosion zone.

We have chosen a setup following that in Leismann et al.\ (2005) which is very similar to that in Komissarov (1999): a cylinder with high pressure ($p_{c}=1$), density ($\rho_{c}=10^{-2}$), and radius $0.8$ is located in the center of a square Cartesian grid which initially contains a uniform, strong magnetic field. Between a radius of $0.8$ to $1.0$ the density and pressure smoothly decrease to those of a homogeneous ambient medium ($\rho =10^{-4}$ and $p=5 \times 10^{-4}$). Initially, the magnetic field is in the $x-$direction, $B_{x}=0.1$, and the velocity is zero everywhere. The simulations are carried out until $t=4.0$ on grid resolutions of $400^2$ and $800^2$, spanning a region of $[-6, 6]^{2}$.

Figure 15 shows the results from the 2D cylindrical explosion test problem at $t=4.0$. 
\begin{figure}[h!]
\epsscale{0.75}
\plotone{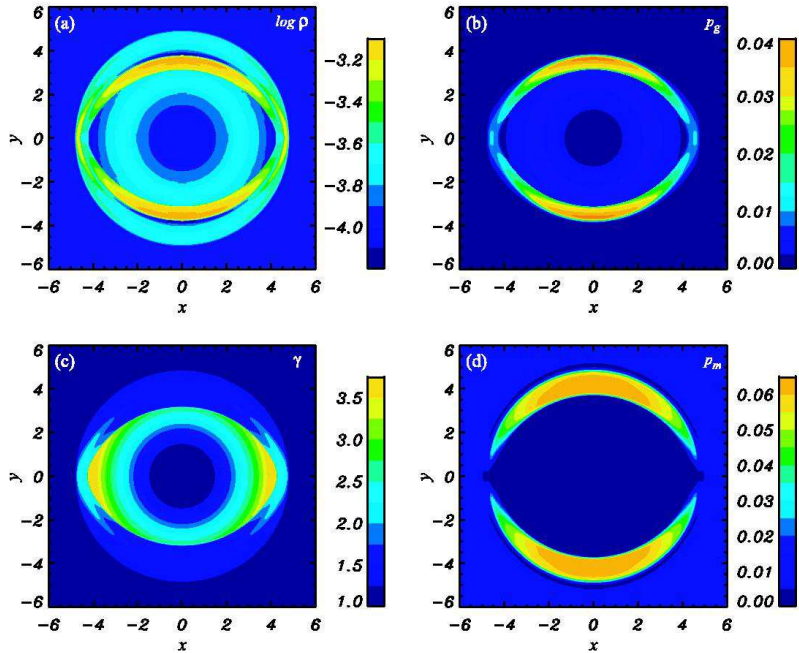}
\caption{2D images of ({\it a}) density, ({\it b}) gas pressure, ({\it c}) Lorentz factor, and ({\it d}) magnetic pressure ($p_{m}=B^{2}/2$) of the 2D cylindrical explosion problem at $t=4.0$ with $400^2$ resolution.
 \label{fap3}}
\end{figure}
The outer fast shock has an almost circular shape. The innermost region is also almost circular and bounded by a reverse fast shock. Between these two fast shocks there are two density shells bounded on the outside by compressed magnetic field. Our test results are qualitatively identical to the results found by other second-order RMHD codes (see Figs. B3 \& B4 in Leismann et al.\ 2005; see Fig.\ 4 in Del Zanna et al.\ 2008).

In order to evaluate the quantitative difference between our simulation code and others, one-dimensional gas pressure and magnetic pressure profiles along the $y$ axis are shown in Figure 16. The location of discontinuities is identical to the results in Leismann et al.\ (2005) and Del Zanna et al.\ (2007). The maximum value of the gas pressure is slightly larger than in Del Zanna et al.\  because of different numerical resolution but the same as in Leismann et al.\ at comparable numerical resolution. Slight differences in the one-dimensional magnetic pressure profile around the maximum result from differences in numerical diffusivity in the different codes.  
\begin{figure}[h!]
\epsscale{0.7}
\plotone{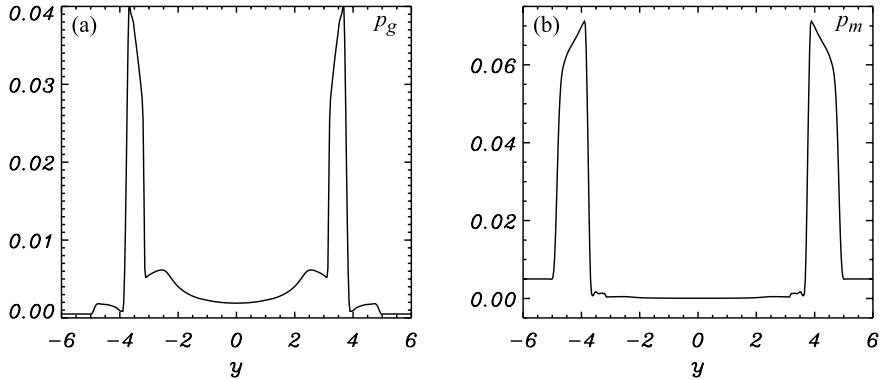}
\caption{1D gas pressure ($left$) and magnetic pressure ($right$) profiles of the 2D cylindrical explosion problem at $t=4.0$ along the $y$ axis using the MC scheme with $400^2$ resolution. 
 \label{fap4}}
\end{figure}

In summary, we obtain qualitatively identical and quantitatively almost identical results when compared to previous second order RMHD simulation code results. We conclude that our simulation code passes this test problem successfully. 

\subsection{3D Rotor}

Del Zanna et al.\ (2003) adapted the two-dimensional rotor problem from non-relativistic MHD performed by Balsara \& Spicer (1999) and T\'{o}th (2000) to the relativistic case. Here we consider a three dimensional version of the relativistic rotor problem (Mignone et al.\ 2009). In a successful test the complicated pattern of shocks and torsional Alfv\'{e}n waves launched by the rotor is handled correctly.

The initial condition consists of a sphere with radius $r_{sp}=0.1$ centered at the origin of the computational box taken to be the unit cube $[-0.5, 0.5]^{3}$. The sphere is heavier ($\rho_{sp} =10$) than the surrounding medium ($\rho=1$) and rapidly rotates around the $z$ axis with an angular velocity $\omega_{sp}=9.95$, i.e., with velocity components $(v_{x}, v_{y}, v_{z})=\omega_{sp}(-y,x,0)$. The gas pressure,  magnetic field and adiabatic index are constant everywhere, $p=1$, $\mathbf{B}=(1,0,0)$, and $\Gamma=5/3$. Exploiting the point symmetry, we have carried out simulations until $t=0.4$ at grid resolutions of $128^3$ and $256^3$. This test problem has resolution-dependent complexity since the maximum Lorentz factor in the initial set up depends on the grid resolution.

Figure 16 shows two-dimensional density images in the $xy$ plane at $z=0$ (perpendicular to the rotation axis) and the $xz$ plane at $y=0$ (including the rotation axis) for the $256^3$ grid resolution case, at $t=0.4$.
When the sphere starts rotating, torsional Alfv\'en waves propagate outwards. 
\begin{figure}[h!]
\epsscale{0.8}
\plotone{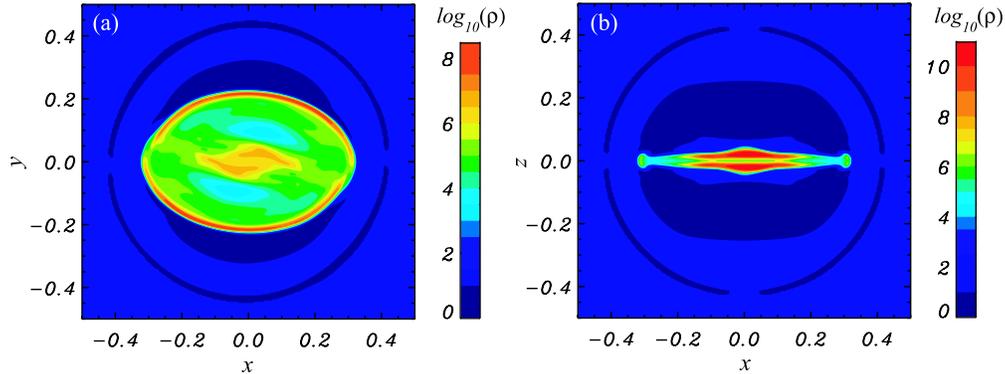}
\caption{2D density image of the 3D rotor problem at $t=0.4$ ({\it a}) in the $xy$ plane at $z=0$  and ({\it b}) in the $xz$ plane at $y=0$  using  the MC scheme with $256^3$ resolution. 
 \label{fap5}}
\end{figure}
The initial spherical structure collapses into a disk-like structure in the equatorial plane ($z=0$) which generates shock waves propagating the vertical direction that can be seen in Figure 16. In the $xy$ plane, surrounding matter is pushed into a thin elliptical shell enclosed by a tangential discontinuity. In Mignone et al.\ (2009) the thin shell shows a distinct octagonal-like shape when they use a five wave HLLD Riemann solver scheme to calculate numerical fluxes that makes a less diffusive transition at rotational (Alfv\'{e}n) discontinuities. The whole structure is embedded in a radially expanding spherical fast rarefaction wave front. In the $xz$ plane our spindle shaped structure appears nearly identical to the results shown in Mignone et al.\ (2009).

Our simulation test results are very similar to the results using an HLL approximate Riemann solver scheme performed by Mignone et al.\ (2009) (see the lower panels of Fig.\ 12 in Mignone et al.\ 2009) except for their now smoother, when compared to the HLLD result, octagonal thin shell structure.  We note that other previous relativistic and non-relativistic results for the MHD 2D rotor test problem using the HLL scheme (Del Zanna et al.\ 2003; Balsara \& Spicer 1999) found a thin elliptical shell structure similar to our result in the xy plane (see the top-left panel in Fig.\ 5 of Del Zanna et al.\ 2003). Therefore the thin elliptical shell structure appears commonly developed in the rotor test problem if the HLL scheme is employed.  Here we must speculate that the difference between the thin elliptical shell structure found by us and others using the HLL scheme, and the smoothed octagonal shell structure found by Mignone et al.\ (2009) using the HLL scheme is associated with a difference in the numerical diffusivity resulting from details of the piecewise linear reconstruction and constrained transport schemes used by Mignone et al.\ (2009). 

Figure 18 shows one-dimensional density profiles along the $y$ and $z$ axes for the different resolutions at time $t=0.4$. 
\begin{figure}[h!]
\epsscale{0.7}
\plotone{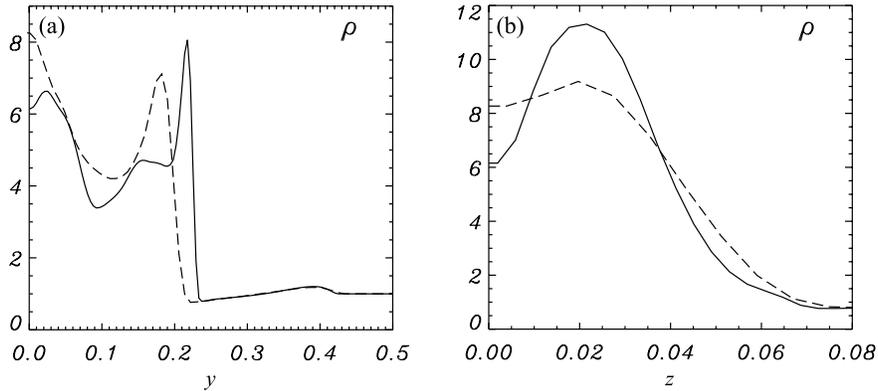}
\caption{\footnotesize \baselineskip 10pt 1D cuts of the 3D rotor problem at $t=0.4$ along the $y$ axis in the equatorial plane (left) and the $z$ rotational axis (right) with $256^3$ resolution (solid) and with $128^3$ resolution (dashed). 
 \label{fap6}}
\vspace{-0.2cm}
\end{figure}
In profiles along the $y$ axis, the density in the central region is larger at lower resolution. Lower resolution gives a lower shell density and underestimates the propagation speed. 
Our simulation test results are very similar to those found using the  HLL approximate Riemann solver scheme in Mignone et al.\ (2009), see the plus and star symbols of Fig.\ 14 in Mignone et al.\ (2009), except for the position of the thin shell. This difference is caused by the difference between the octagonal shape of the thin shell found by  Mignone et al.\ and the elliptical shape of the thin shell found here. A protrusion associated with the octagonal-like thin shell along the $y$ axis extends beyond the smoother elliptical shell. 

In summary, we obtain very similar results to previous studies of the 3D and 2D rotor test problem using the HLL approximate Riemann solver scheme in RMHD. We conclude that our simulation code passes this test problem as successfully as other second-order codes.

\end{document}